\documentclass[12pt, notitlepage]{article}
\pdfoutput=1

\usepackage[toc,page]{appendix}

\usepackage{float}
\usepackage{subfig}
\usepackage{graphicx}
\usepackage[T1]{fontenc}
\usepackage[utf8]{inputenc}
\usepackage{authblk}
\usepackage{amsmath,bm}
\usepackage{mathtools}
\usepackage[left=3cm,right=3cm,top=2cm,bottom=2cm]{geometry}
\renewcommand\footnotemark{}

\usepackage{color}

\newcommand{\rf}[1]{(\ref{#1})}
\newcommand{\beq}{\begin{equation}}
\newcommand{\beql}[1]{\beq\label{#1}}
\newcommand{\eeq}{\end{equation}}
\newcommand{\bea}{\begin{eqnarray}}
\newcommand{\eea}{\end{eqnarray}}

\newcommand{\eps}{\epsilon}

\begin{document}

\title{The phase structure of Causal Dynamical Triangulations with toroidal spatial topology}         

\author[a,b]{J.~Ambj\o rn}
\author[c]{J.~Gizbert-Studnicki}
\author[c]{A.~G\"orlich}
\author[c]{J.~Jurkiewicz}
\author[c,d]{D.~N\'emeth}
\affil[a]{\small{The Niels Bohr Institute, Copenhagen University, \authorcr Blegdamsvej 17, DK-2100 Copenhagen Ø, Denmark. \authorcr E-mail: ambjorn@nbi.dk.\vspace{+2ex}}} 

\affil[b]{\small{IMAPP, Radboud University, \authorcr Nijmegen, PO Box 9010, The Netherlands.\vspace{+2ex}}}

\affil[c]{\small{The M. Smoluchowski Institute of Physics, Jagiellonian University, \authorcr \L ojasiewicza 11, Krak\'ow, PL 30-348, Poland. \authorcr Email: jakub.gizbert-studnicki@uj.edu.pl, andrzej.goerlich@uj.edu.pl, jerzy.jurkiewicz@uj.edu.pl.\vspace{+2ex}}}

\affil[d]{\small{E\"otv\"os Lor\'and University,   \authorcr P\'azm\'any P\'eter s\'et\'any 1/A, 1117 Budapest, Hungary.
\authorcr Email: kregnach@caesar.elte.hu.}}

\date{\small({Dated: \today})}          
\maketitle


\begin{abstract}
We investigate the impact of topology on the phase  structure of four-dimensional Causal Dynamical Triangulations (CDT).  Using numerical Monte Carlo simulations we study CDT with toroidal spatial topology. We confirm  existence of all four distinct phases of quantum geometry earlier observed in CDT with spherical spatial topology. We plot the toroidal CDT phase diagram and {find that  it looks very similar to the case of the spherical spatial topology.}

\vspace{1cm}

\end{abstract}


\begin{section}{Introduction}\label{intro}

Asymptotic safety provides an exciting possibility of formulating a  nonperturbative and background independent theory of Quantum Gravity. The idea was put forward in a seminal paper of  S. Weinberg ~\cite{Weinberg79} who proposed to extend the notion of renormalizability into {the} nonperturbative regime defined at non-Gaussian ultraviolet fixed point(s) (UVFP) of renormalization group trajectories.  If   {the}  renormalization group flow originating in the UVFP lies  (in an abstract space of coupling constants) on a  hypersurface of finite dimension then only a finite number of (important) couplings are attracted to the UVFP. Such couplings could  in principle {be} determined by a finite number of experiments making the whole formulation finite and predictive for arbitrarily large energy scale. As a result it might be possible to overcome the well-known problem of (perturbative) nonrenormalizability of gravity treated as a conventional quantum field theory (QFT) expanded around any fixed background geometry~\cite{Goroff:1985th,'tHooft:1974bx}. There are known examples of perturbatively nonrenormalizable but asymptotically safe QFTs \cite{Brezin:1976, Gawdzki:1985,Wellegehausen:2014noa}, and for gravity the Weinberg's conjecture is strongly supported by functional renormalization group studies \cite{Reuter:1998,Litim:2003vp, Niedermaier2006, Codello:2008vh,Benedetti:2009rx,Litim:2011cp} which provide growing evidence that gravitational UVFPs really exist. 
A key difficulty remains:  in order to investigate an asymptotically safe theory in full-glory one is forced to apply nonperturbative tools.
Such tools are provided in the research program of Causal Dynamical Triangulations (CDT) \cite{Ambjorn:1998xu} which is an attempt to quantize gravity based on a latice regularization of {the} nonperturbative path integral  
\begin{equation}\label{PI}
{\cal Z}_{QG} =\int {\cal D}_{\cal M}[g] \   e^{iS_{HE}[g]} 
\end{equation}
over spacetime geometries $[g]$, i.e. equivalence classes of  metrics $g$ with respect to
diffeomorphisms, and $S_{HE}$ is {the} classical Einstein-Hilbert action.
To give precise meaning to expression (\ref{PI}) CDT introduces {a} lattice regularization by constructing (continuous) spacetime geometries from four-dimensional  simplicial building blocks.
 It is assumed that spacetime inside each such block is flat  and {the} nontrivial geometry 
 depends on how these blocks are glued together 
 (e.g. {a} local deficit angle in {D dimensions} is encoded in {the} number of simplices sharing a given D-2 dimensional subsimplex). By gluing simplices together one obtains
piecewise linear simplicial manifolds, also called triangulations, and the formal path integral  (\ref{PI}) is defined as
\begin{equation}\label{Z}
{\cal Z}_{CDT} \equiv \sum_{\cal T}e^{i S_{R}[{\cal T} ]} \ ,
\end{equation}
where the sum is over triangulations  ${\cal T}$, and $S_R$ is the Einstein-Hilbert action for a triangulation obtained following Regge's method for describing piecewise linear simplicial geometries  \cite{Regge:1961px}. 

A theory of quantum gravity should describe {spacetime} at {the} Planck scale, where one may expect large fluctuations of {the} geometry. Such fluctuations could in principle lead to changes in spatial topology. Therefore there is a longstanding discussion if in {the} path integral (\ref{Z}) one should allow for topology changes and, if so, which topologies should be taken into account. Including topological fluctuations was considered in the easiest case of two-dimensional toy {models} in the Euclidean formulation \cite{Francesco:1995}. It was shown that in this case a naive understanding of the topology of the Universe breaks down and if we want to stay in a formalism similar to QFT we must suppress such fluctuations. The problem is that there are many more geometries of complicated topology than there are of simple topology and any sum over geometries is dominated by these complicated topologies and plainly divergent.\footnote{The number of geometric configurations of a given genus $h$ grows super-exponentially with $h$ and the resulting genus expansion of the path integral {is} not even Borel-summable.} The problem is somehow eased in {the} Lorentzian formulation  where one is able to define  two-dimensional {models} with spatial topology fluctuations but at a cost of restricting the class a geometries taken into account to those causally 'well-behaved' \cite{Loll2003,Loll2006,Loll2005,Loll2006AIP}. In higher dimensions the situation becomes much worse, and in particular in four dimensions the problem is not even well posed as four-dimensional topologies are not classifiable. Therefore in CDT one adopts a pragmatic point of view by considering only {spacetimes} admitting 
{a} global proper time $T$ foliation into spatial hypersurfaces $\Sigma$ of fixed topology.
 This requirement is compatible with imposing {an} additional causal structure of global hypebolicity on admissible geometries (triangulations) which enter { the} path integral (\ref{Z}), such that each triangulation ${\cal T}$ is topologically ${\cal M} =\Sigma \times T$.  {The idea of an imposed global time-foliation also appears in  Ho\v{r}ava-Lifshitz gravity \cite{Horava:2009uw}, which is an attempt to define the UV completion of general relativity by introducing anisotropic scaling of space-time coordinates in the high-energy regime. It has been actually shown that the continuum limit of two-dimensional CDT is compatible with the two-dimensional projectable Ho\v{r}ava-Lifshitz gravity \cite{Ambjorn:2013joa,Glaser:2016PRD}, and that both approaches share many features in three \cite{Benedetti:2009ge,Anderson:2011bj,Budd:2012,Benedetti:2015} and four \cite {Horava:2009if, Ambjorn:2010hu, Mielczarek:2017b} space-time dimensions.} 
 
 In four-dimensional CDT, each spatial layer of integer (lattice) time $t$ is constructed from equilateral tetrahedra with space-like links $a_s$. The neighbouring   spatial layers at $t$ and $t\pm 1$ are linked by four-(dimesional-)simplices, with additional time-like links $a_t$, glued together in such a way that also all intermediate spatial layers  between $t$ and $t\pm1$ have  the requested fixed topology $\Sigma$. One can show that the four-dimensional simplicial complex obeying {the} CDT topological restrictions can be constructed from just two types of building blocks (see Fig. \ref{F:Simplices}), the $(4,1)$ simplex with $4$ vertices (a tetrahedron) in time $t$ and one vertex in $t\pm1$, and the $(3,2)$ simplex with $3$ vertices (a triangle) in $t$ and $2$ vertices (a link) in $t\pm 1$. In this case the Einstein-Hilbert-Regge action takes a form \cite{Ambjorn:2001cv}
\beql{SRegge}
S_{R}=-\left(\kappa_{0}+6\Delta\right)N_{0}+\kappa_{4}\left(N_{(4,1)}+N_{(3,2)}\right)+\Delta \ N_{(4,1)} \  ,
\eeq 
where $ N_{(4,1)}$,  $ N_{(3,2)}$ and $N_0$ denote the total number of $(4,1)$ simplices, $(3,2)$ simplices and vertices, respectively. The  action includes three bare dimensionless coupling constants $\kappa_{0}$, $\Delta$ and $\kappa_{4}$. $\kappa_{0}$ is inversely proportional to the bare Newton's constant, $\Delta$ is related to the ratio of the length of space-like and time-like links on the lattice  ($a_t^2 = - \alpha\  a_s^2$, where $\alpha > 0$ is called the asymmetry  parameter) and $\kappa_{4}$ is proportional to the bare cosmological constant. 

\begin{figure}[H]
\centering
\includegraphics[width=0.42\linewidth]{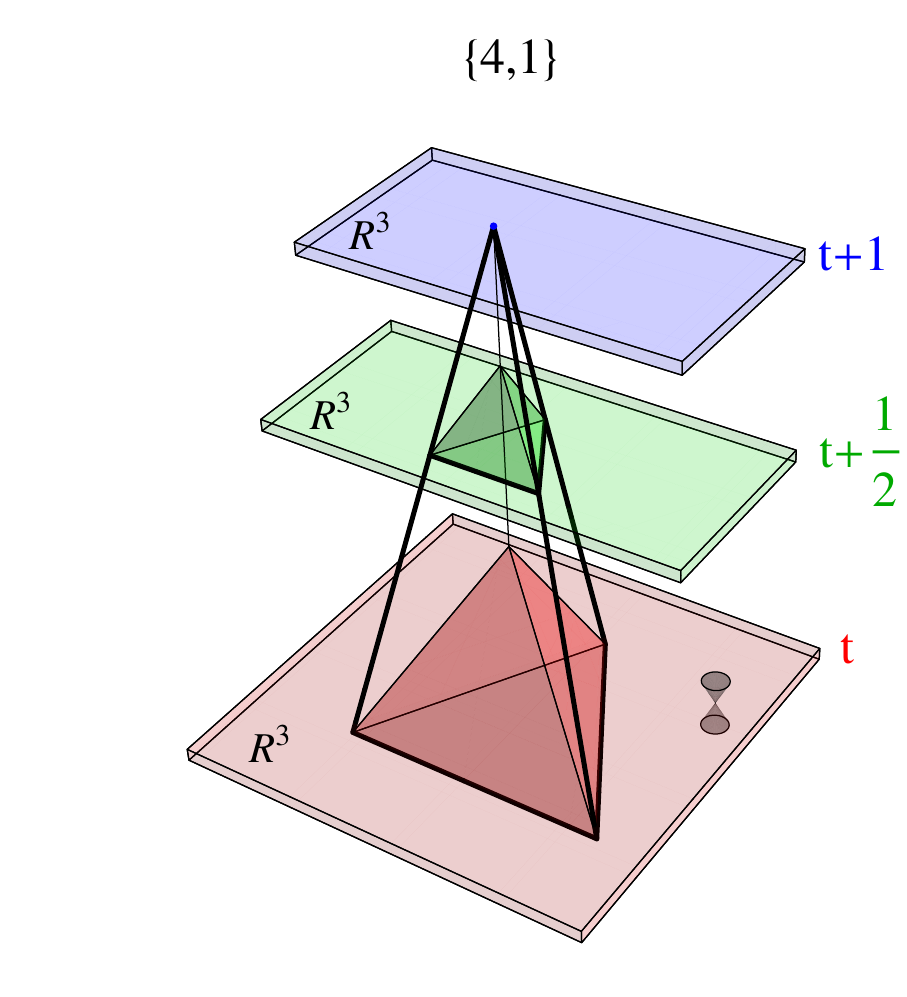}
\includegraphics[width=0.42\linewidth]{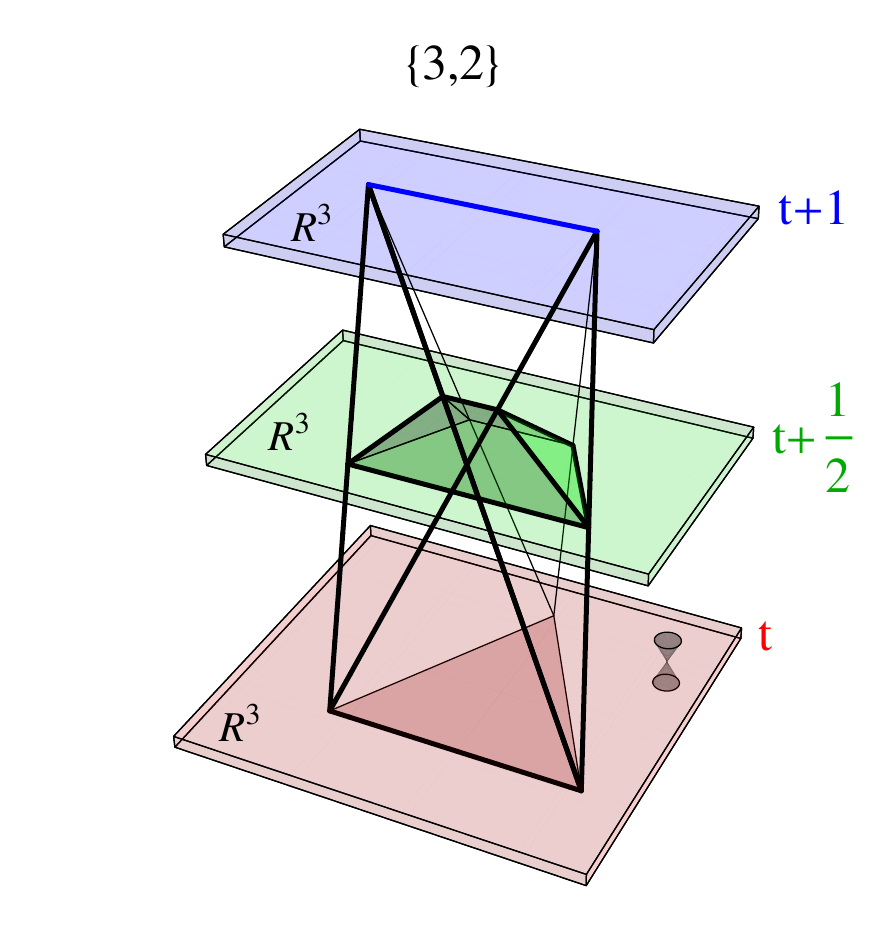}
\caption{\small {Visualization of the elementary building blocks of four-dimesnional CDT, the $(4,1)$-simplex (left) and the $(3,2)$-simplex (right). }}
\label{F:Simplices}
\end{figure}

Such a formulation is coordinate-free (all geometric degrees of freedom are expressed by  
  topological invariants - lengths and angles) and nonperturbative (all possible triangulations are included in the path integral (\ref{Z})). It is also manifestly background independent as there is no preferred  triangulation put in by hand but macroscopic  geometry emerges dynamically from fluctuations.
  It is important to note that no ad hoc discreteness of spacetime is assumed from the outset, and the discretization appears only as a regularization of the path integral (\ref{PI}). 
{The} finite length {$a$ of the links} in a triangulation constitutes an ultraviolet cutoff   which is intended to be removed in the continuum limit $a \to 0$, which should be consistent with the perspective UVFP of the renormalization group trajectories. In a lattice formulation, as CDT,  the UVFP should appear as a  phase transition  of second or higher order, where  infinite correlation lengths enable one to go simultaneously with {the} lattice spacing  $a \to 0 $ and {the linear lattice size $L \to \infty$, such that physical lengths $a L$} remain constant and thus  observable quantities expressed in physical units are kept fixed. Therefore  analysis of {the} CDT phase structure   and {the} order of {the} phase transitions constitute   first steps in a quest for the ultraviolet limit (in   CDT formulation) of quantum gravity. At the same time it is very important  to be able to correctly reproduce
the infrared limit compatible with classical Einstein's general relativity.

This paper is organised as follows. After reviewing some technical details regarding {the} numerical implementation in section~\ref{Numerics} we summarize the state of the art of CDT in section~\ref{StateOfArt}. In section~\ref{OrderParam} we define {the} order parameters and we explain how to study the CDT phase structure. In section~\ref{PhaseStruc} we present  the results of  numerical simulations performed {with the spatial topology fixed to that  of a three-torus}. A discussion and a summary of  {the} results obtained in this work are presented in section~\ref{Discussion}.  

\end{section}  

\begin{section}{How to perform numerical simulations}\label{Numerics}

The study of the regularized path integral (\ref{Z}) in four spacetime dimensions requires using numerical methods. This is possible by  applying {a} Wick rotation of the proper time coordinate from Lorentzian (real) time $t^{(L)}$ to Euclidean (imaginary) time  $t^{(E)}~=~-it^{(L)}$. Such a Wick rotation is well defined in CDT due to {the} assumed spacetime foliation into Cauchy hypersurfaces of constant proper time. The causal structure enables one to change time-like links into space-like links by changing the asymmetry parameter  $\alpha \to -\alpha$ and  accordingly the length of the time-like links becomes $a_t^2 =  \alpha\  a_s^2 \ , \ \alpha > 0 $. As a result the four-simplices become parts { of  Euclidean space}. At the same time one should change {the} Lorentzian action $S_R^{(L)}$ into {the} Euclidean action  $S_R^{(E)}=-iS_R^{(L)}$ by analytically  continuing the bare coupling constants $\kappa_0$, $\Delta$ and $\kappa_4$ from eq.~(\ref{SRegge}), which are all analytic functions of $\alpha$ in the lower half of the complex $\alpha$ plane.\footnote{The functional form of the bare Euclidean Einstein-Hilbert-Regge action $S_R^{(E)}$ is the same as for the Lorentzian action $S_R^{(L)}$ from eq. (\ref{SRegge}). Of course the form {of   $\kappa_0$, $\Delta$ and $\kappa_4$ as functions} of $\alpha$ is modified, but anyway they are bare coupling constants and as such their functional dependence of the bare Newton's constant, cosmological constant and $\alpha$ is irrelevant.} Consequently the path integral of CDT~(\ref{Z}) becomes a partition function of a statistical theory of random geometries
\beql{ZE}
{\cal Z}_{CDT} = \sum_{{\cal T}^{(L)}}e^{i S_{R}^{(L)}[{\cal T} ^{(L)}]} \ \xrightarrow[\alpha \, \to \, -\alpha]{} {\cal Z}_{CDT} = \sum_{{\cal T}^{(E)}}e^{-S_{R}^{(E)}[{\cal T}^{(E)} ]} \ .
\eeq
One should keep in mind that the class of admissible  (Euclidean) triangulations ${\cal T}^{(E)}$ which enter the partition function (\ref{ZE}) keeps track of the imposed Lorentzian structure by the  proper time foliation constraint, and it is not the same as for Euclidean Dynamical Triangulations (EDT) where such a constraint is absent. EDT have not been able to correctly reproduce the suitable infrared limit  \cite{Ambjorn:1991pq, Agishtein:1991cv, Catterall:1994pg,Ambjorn:2013eha, Coumbe:2014nea} nor to define a  continuum limit\footnote{Authors of Ref. \cite{Laiho:2016nlp} conjecture that a continuum limit may exist if bare coupling constants are fine-tuned in a specific way.} \cite{Bialas:1996wu, deBakker:1996zx}, irrespective of the spacetime topology chosen \cite{Bilke:1996zd}. The problem with EDT probably lies in the fact that the formulation is Euclidean from the outset and thus the distinction between space and time is lost and time has to be reconstructed dynamically, which leads to its pathological behaviour. These problem seems to be cured in CDT, where time is treated semi-classically.

The above construction makes it possible to  study the partition function (\ref{ZE})  using numerical Monte Carlo techniques. One starts from an arbitrary simple initial triangulation (configuration of simplices), consistent with the requested fixed spacetime topology, and updates its geometry   using Monte Carlo moves. The moves are local (they create, delete or reconstruct simplices in the closest neighbourhood of some randomly chosen place in a triangulation),  causal (preserve fixed spacetime topology {as well as the foliation structure}) and  ergodic (any triangulation obeying CTD topological restrictions  is achievable from any other triangulation by a sequence of moves). By using {the} Metropolis algorithm the numerical code applies the moves in such a way that the system performs random walk in the space of admissible triangulations and, after a thermalization period, statistically independent triangulations ${\cal T}^{(E)}$ are generated  with probabilities consistent with   Boltzman weights $\propto \exp({-S_{R}^{(E)}[{\cal T}^{(E)}]})$. As a result one can use generated triangulations to estimate expectation values or correlators of  observables.

In a typical Monte Carlo simulation one sets the values of the bare coupling constants: $\kappa_0$, $\Delta$ and $\kappa_4$. Simulations show that for fixed values of $\kappa_0$ and $\Delta$ the {number  of  triangulations with 
a fixed lattice volume  $N_4 = N_{(4,1)}+ N_{(3,2)}$ is, to leading order, proportional to $ \exp(\kappa_4^c N_4)$, i.e. it grows approximately exponentially with $N_4$.  $\kappa_4^c$ is a function of  $\kappa_0$ and $\Delta$}. Due to this entropic effect, and thanks to the fact that  in the bare CDT action (\ref{SRegge}) the $\kappa_4$ is conjugate to $N_4$, the leading behaviour of the partition function (\ref{ZE}) is  $ {\cal Z} (\kappa_0, \Delta ,\kappa_4) \propto \exp ((\kappa_4^c-\kappa_4) N_4$) and { the partition function} is divergent for $\kappa_4 < \kappa_4^c$. In numerical simulations it is  more practical to {fix}  the total lattice volume $N_4$ and in such a case one should also {choose} $\kappa_4\approx \kappa_4^c(N_4,\kappa_0, \Delta)$.\footnote{By performing {simulations} this way one in fact investigates the properties of ${\cal Z}(\kappa_0, \Delta ,N_4)$ which is {related to} ${\cal Z}(\kappa_0, \Delta ,\kappa_4)$ by {a Laplace transformation. One can  investigate the infinite volume limit as well as reconstruct  ${\cal Z}(\kappa_0, \Delta ,\kappa_4)$ from extrapolations of 
measurements done for  different   $N_4$.}} The volume is usually {controlled}  by introducing an additional volume fixing term to the bare action. In this work we use {a} quadratic volume fixing \footnote{Note that technically we fix $N_{(4,1)}$ instead of $N_4$, but for fixed values of $\kappa_0$ and $\Delta$ the ratio $N_4 / N_{(4,1)}$ is approximately constant and volume fixing method does not have much impact on the results.}
\beql{Vfix}
S_{VF}=\eps( N_{(4,1)} -  \bar N_{(4,1)})^2 \ ,
\eeq
which makes the total number of $(4,1)$ simplices oscillate around $\bar N_{(4,1)}$, and the impact of such volume fixing can { easily be } removed from the numerical data.

Finally, before starting {the} numerical simulations, one should choose the fixed topology  of spatial slices $\Sigma$ and  the length / boundary conditions for the proper time axis $T$. For practical reasons the CDT simulations are usually done for a periodic time axis of fixed length with $t_{tot}$ spatial slices, resulting in a global spacetime topology ${\cal M} = \Sigma \times S^1$. Most of the previous results of CDT  were obtained for the spatial topology of a three-sphere $\Sigma = S^3$, and in this article we focus on  the spatial topology of a three-torus $\Sigma = T^3 = S^1 \times S^1 \times S^1$. 

Summing up, in the numerical simulations described below we set: ${\cal M} = T^3 \times S^1$, $t_{tot} = 40 \text{ or } 4$, $\eps = 0.00002$, $\bar N_{(4,1)} = 80000  \text{ or } 160000$ and we scan the parameter space spanned by $\kappa_0$ and  $\Delta$. For each data point we adjust  $\kappa_4$ to the critical value $ \kappa_4^c(\bar N_{(4,1)},\kappa_0, \Delta)$, see Fig. \ref{F:K4}, so we are effectively left with  a two-dimensional parameter space: ($\kappa_0, \Delta$).

\begin{figure}[H]
\centering
\includegraphics[width=0.65\linewidth]{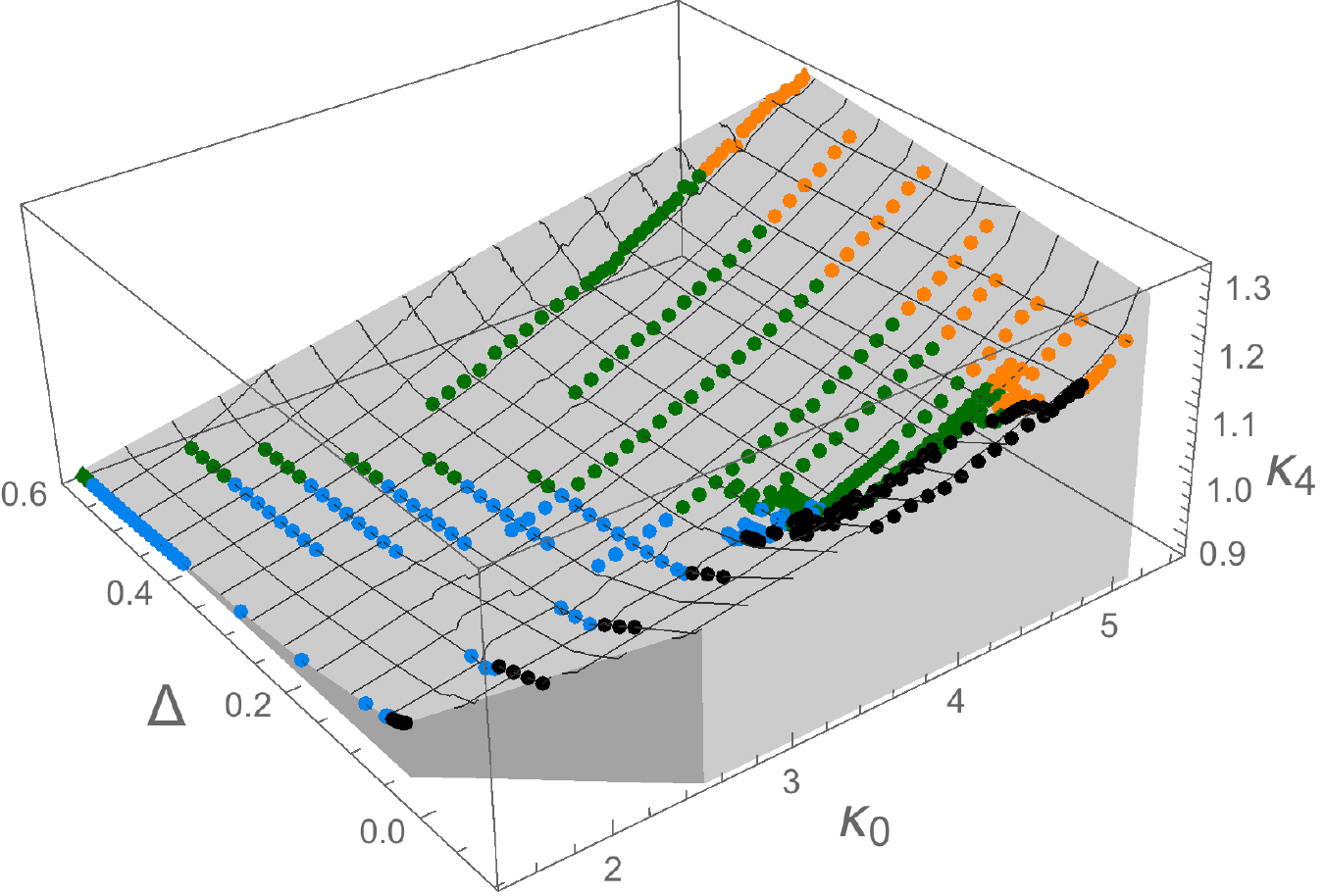}
\caption{\small {The critical surface $\kappa_4^c(\kappa_0, \Delta)$ for fixed $\bar N_{(4,1)}=160000$ and $t_{tot} = 4$, measured in spacetime topology ${\cal M} = T^3 \times S^1$. Colors denote various phases of geometry described in Section \ref{PhaseStruc}.}}
\label{F:K4}
\end{figure}

\end{section}


\begin{section}{State of the art}\label{StateOfArt}
The key assumption of CDT is a choice of fixed spatial topology used in computer simulations. Most of the numerical studies performed in the past were for {the}  specific choice of a three-sphere and time-periodic boundary conditions resulting in the global spacetime topology ${\cal M} = S^3 \times S^1$. This particular choice led to many interesting results, including discovery of four distinct phases of spacetime geometry, historically called $A$, $B$, $C_{dS}$ \cite{Ambjorn:2005qt,Ambjorn:2010hu} and $C_b$ \cite{Ambjorn:2014mra,Ambjorn:2015qja}, see Fig. \ref{F:phaseold}. 

\begin{figure}[H]
\centering
\includegraphics[width=0.65\linewidth]{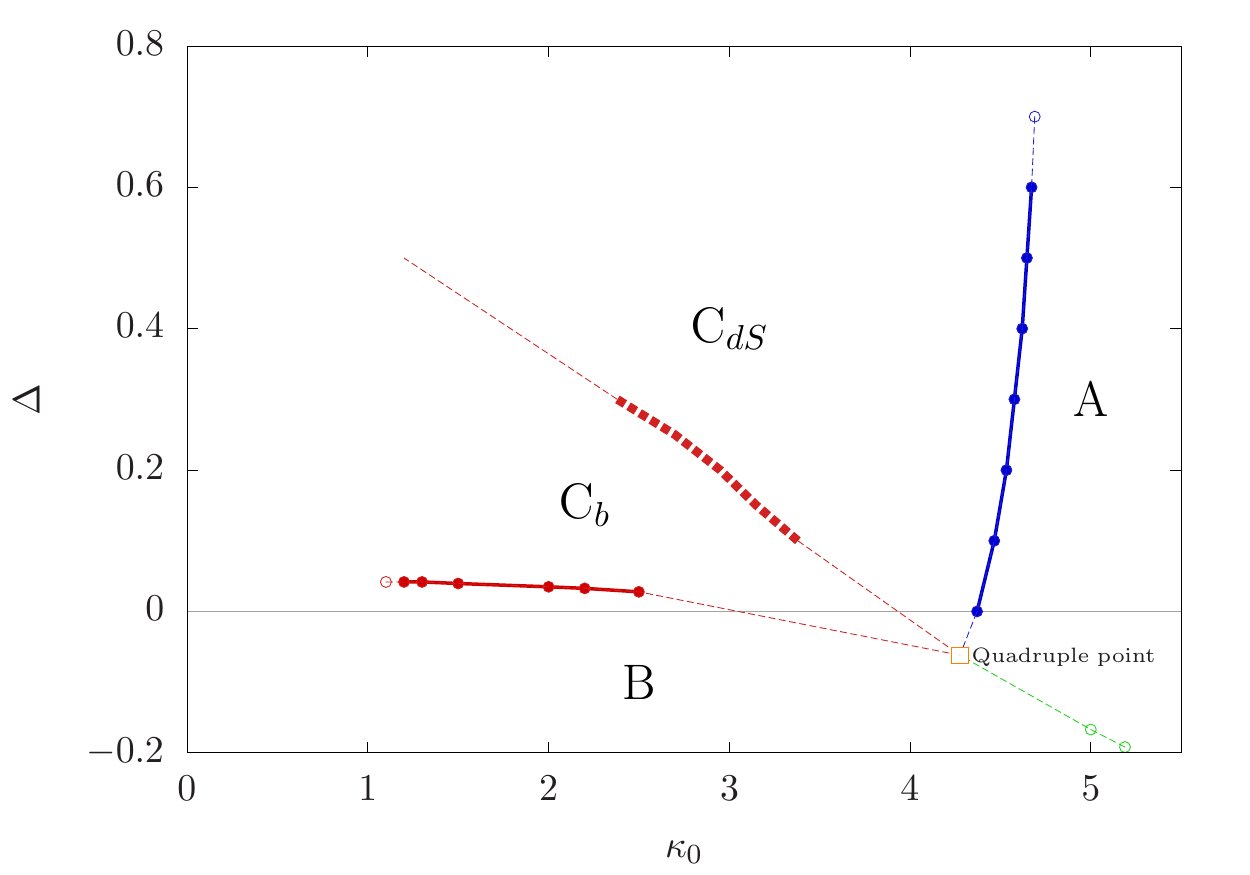}
\caption{\small {Phase diagram of CDT with spherical spatial topology.}}
\label{F:phaseold}
\end{figure}

\begin{figure}[H]
\centering
\includegraphics[width=0.32\linewidth]{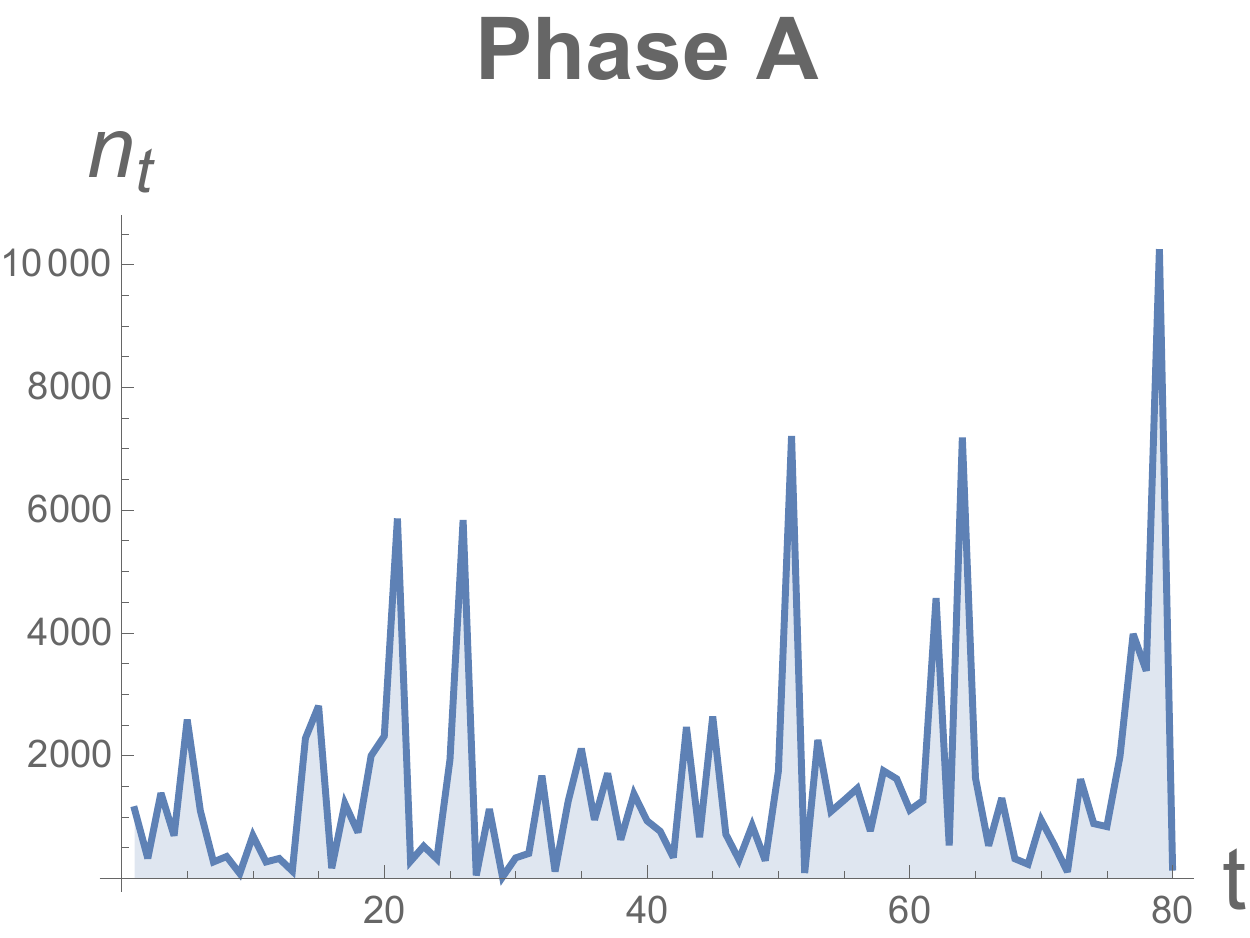}
\includegraphics[width=0.32\linewidth]{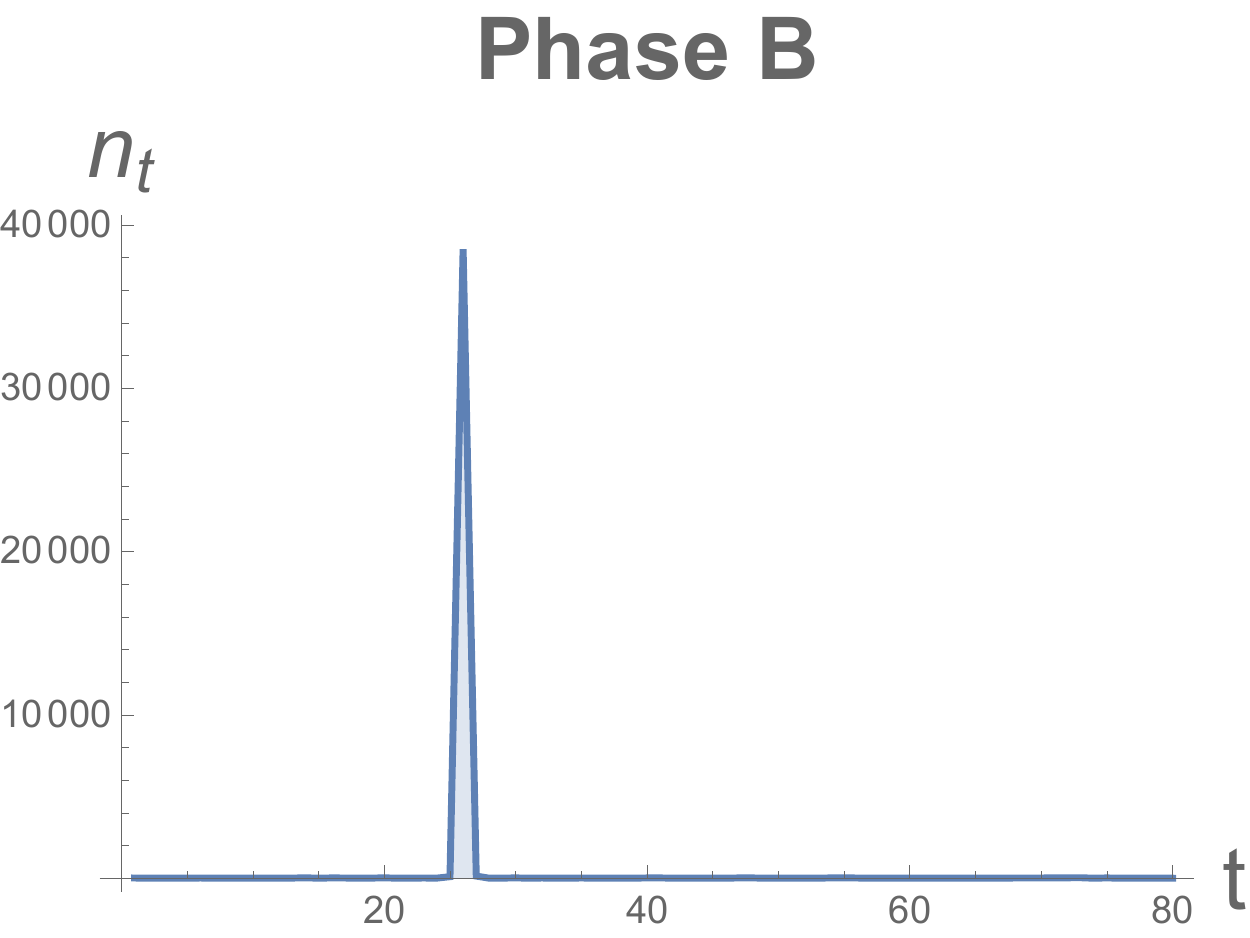}
\includegraphics[width=0.32\linewidth]{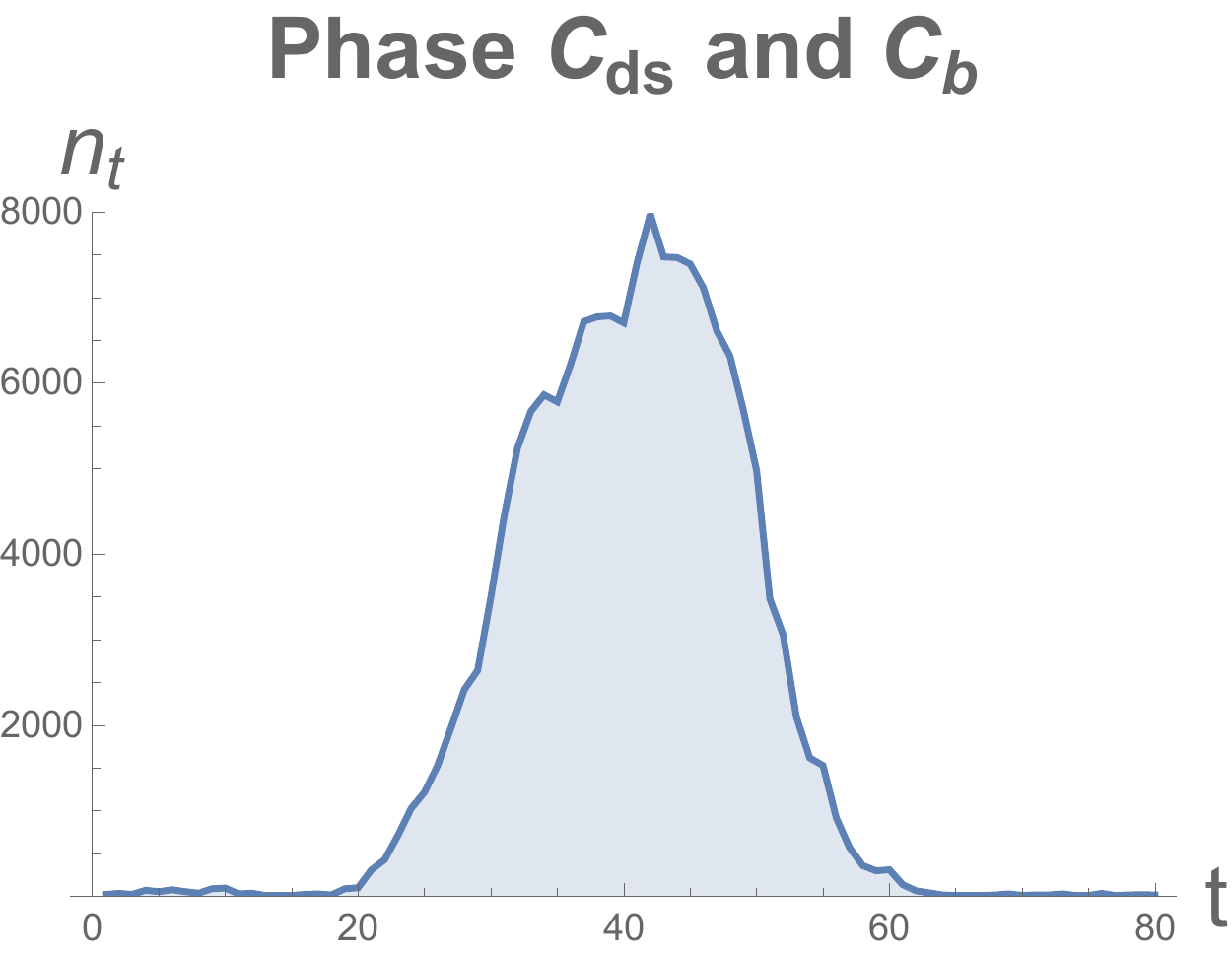}
\caption{\small {Typical volume profiles $n_t \equiv N_{(4,1)}(t)$ in various CDT phases in spherical spatial topology $\Sigma=S^3$.}}
\label{F:VolSphere}
\end{figure}

In phase $A$ spacetime disintegrates into many causally uncorrelated baby universes with very short proper time extension, see Fig. \ref{F:VolSphere} where the typical spatial volume profile $n_t \equiv N_{(4,1)}(t)$ (i.e. the number of simplices with 4 vertices in lattice time $t$) is plotted.\footnote{For historical reasons we keep a convention in which $n_t$ is in fact twice the volume of a spatial slice $t$, i.e. twice the number of equilateral spatial tetrahedra forming a spatial slice $t$.}  The Hausdorff and spectral dimensions of triangulations inside this phase are approximately  equal two. The phase is the CDT analogue of the branched polymer phase observed earlier in EDT. In phase $B$ the spacetime geometry collapses into a single spatial slice, see Fig. \ref{F:VolSphere},  with  (probably) infinite  Hausdorff and  spectral dimensions. The phase is the CDT analogue of the crumpled phase of EDT and it does not have a physical interpretation. A  nontrivial  result is the existence of the phase $C_{dS}$, also called the de Sitter phase, where one observes  {the} dynamical emergence of {a} large scale four-dimensional geometry  \cite{Ambjorn:2005qt,Ambjorn:2004qm}  consistent with a semiclassical (Euclidean) de Sitter universe \cite{Ambjorn:2007jv,Ambjorn:2008wc}, see Fig. \ref{F:VolSphere}. At the same time {the} spectral dimension shows 
{a} non-trivial scale dependence and ranges from $\approx 2$ for short scales to $4$ at large scales (diffusion times) \cite{Ambjorn:2005db, Coumbe2015, Mielczarek:2017}. In the phase $C_{dS}$ {the  quantum fluctuations of the spatial volume} are well described by the minisuperspace reduction of the Hilbert-Einstein action \cite{Ambjorn:2008wc,Ambjorn:2011ph}, and the contribution of such fluctuations vanishes in the infinite volume limit. This phase can be interpreted as the infrared limit of CDT, consistent with Einstein's GR.
The recently discovered phase $C_b$, also called the bifurcation phase \cite{Ambjorn:2014mra}, has a very nontrivial spacetime geometry. The volume profile $n_t$  resembles the one observed in the $C_{dS}$ phase, see Fig. \ref{F:VolSphere}, but   spatial homogeneity is strongly broken by the appearance of compact spatial volume clusters concentrated around vertices with macroscopically large coordination numbers present in the every second spatial layer \cite{Ambjorn:2015qja,Ambjorn:2017}. This phase is still being studied carefully and conclusive physical interpretation of its geometry has not yet been found.

The  phases are separated by first order ($A-C_{dS}$) and  second (or higher) order ($B-C_b$) \cite{Ambjorn:2011cg,Ambjorn:2012ij} phase transition lines. The recently discovered  $C_{dS} - C_b$ phase transition is also second (or higher) order \cite{Coumbe:2015oaa,Ambjorn:2017nho} and all the phases {might} meet in a common  point.\footnote{The existence of such a common 'quadruple' point is {entirely conjectual (and maybe even unlikely) and just based on 
a not too precise  extrapolation} of the measured phase transition lines. Unfortunately our Monte Carlo algorithm looses efficiency in the vicinity of this point, resulting in extremely long autocorrelation times, which {currently} makes simulations in this region of {the} CDT phase space intractable.} This   may  in principle allow one to establish the perspective continuum limit by approaching a second order transition line or a multiple point from  the physically interesting phase $C_{dS}$, thereby defining a smooth interpolation between the low and high energy regimes of CDT \cite{Ambjorn:2014gsa,Ambjorn:2016cpa}.

All  results described above were obtained for a spherical spatial topology $\Sigma=S^3$, but { recently we have been
interested also in imposing toroidal spatial topology. One reason this is of interest is the background independence 
of CDT. No background {\it geometry} is imposed. However, in the case of spherical spatial topology we saw  semiclassical 
four-dimensional (Euclidean) de Sitter-like configurations emerge, around which there were well defined quantum fluctuations.
By changing the spatial topology to  a toroidal  topology $\Sigma=T^3$ one would expect different semiclassical solutions
to dominate, so if the emergence of semiclassical geometry is a universal aspect of CDT one would expect to observe 
completely different geometries at least for choices of bare coupling constants where one
obtained semiclassical configurations  for spherical spatial topology.
Further, we were} inspired by the functional renormalization group research adapted to the ADM-formalism, where the authors of  \cite{Biemans2017,Biemans2017b} started to investigate time foliated (Euclidean) spacetimes  with  topology  $T^d\times S^1$. Recently a similar study was performed for the  topology $S^d \times S^1$ \cite{Houthoff2017} leading to conclusions that the renormalization group flow linking IR and UV fixed points is essentially independent of the  spatial topology chosen.  Our results suggest that in CDT with a fixed spatial topology of a three-torus ($\Sigma = T^3$) there exists a semiclassical phase $C$, similar to phase $C_{dS}$ earlier observed for the topology of a three-sphere ($\Sigma = S^3$) \cite{Ambjorn:2016fbd,Ambjorn:2017226}. However the dynamically generated (flat) background geometry  of the new semiclassical phase $C$ observed in toroidal  topological conditions is completely different than the (four-sphere) geometry of  phase $C_{dS}$ observed for spherical topological conditions. {As mentioned above this is a quite non-trivial result and it gives strong
support to the idea that there is a phase of CDT where semiclassical geometries emerge. We have in addition shown 
that the spatial volume fluctuations are still well described by a suitable minisuperspace reduction of the (toroidal) Einstein-Hilbert action in this phase, and one is even able to measure quantum corrections with much higher precision in the toroidal case.} 

In the current work we want to investigate the existence of other phases in CDT with toroidal  topological  conditions $\Sigma = T^3$ and to check if the phase diagram is similar to the case of spherical topology. As phase transition studies are very resource consuming, in this paper we  focus on the phase structure itself, and the order of the phase transitions will be investigated in forthcoming articles.

\end{section}


\begin{section}{Order parameters for the phase transitions}\label{OrderParam}

Before investigating the phase diagram of CDT one should define appropriate order parameters ($OPs$) that capture  differences between generic triangulations observed in various  phases of  quantum gravity. The differences are usually caused by breaking of some global or local symmetries of the  triangulations making the order parameter jump (first order transition) or rapidly change (second or higher order transition) from one phase to another. At the same time there is some freedom in  choosing the order parameters and one should do it carefully to get  clear  signals of the transitions.

In this work we focus on four order parameters, selected in such a way that they capture  {the} most important differences between {the} CDT phases. We use  {the} experience gained in  previous studies of CDT with spherical spatial topology, where  various phases of geometry were observed  (see Section \ref{StateOfArt}) and similar order parameters were used in the study of the phase transitions. The order parameters discussed in this paper can be divided into two groups. The first group comprises order parameters which capture very global properties of CDT triangulations, such as
\beql{E:OP1}
OP_1 = N_0 / N_{4} \ ,
\eeq   
\beql{E:OP2}
OP_2 = N_{(3,2)} / N_{(4,1)} \ ,
\eeq
which were previously used in the analysis of the $A-C_{dS}$ and $B-C_b$  transitions observed in   
{ the case of} spatial topology $S^3$  \cite{Ambjorn:2011cg,Ambjorn:2012ij}. The $A-C_{dS}$ transition was related to the time-translation symmetry breaking from (a symmetric) phase $A$ to a (less symmetric) phase $C_{dS}$ where one could see a spacetime blob structure, i.e. some macroscopic extension of the universe in time direction. This time-extended universe persisted in phase $C_b$ and the symmetry was further broken in (the least symmetric) phase $B$, where the universe collapsed to a single spatial layer, see Fig. \ref{F:VolSphere}. Such symmetry differences resulted in the order parameter $OP_1$ being large in phase $A$, medium in phases $C_{dS}$ and $C_b$, and small in phase $B$, see Fig. \ref{F:OPsphere}. At the same time the breaking of causal connections of neighbouring spatial layers caused the order parameter $OP_2$ to be small in (time uncorrelated) phases   $A$ and $B$ and large in (time correlated) phases $C_{dS}$ and $C_b$, see Fig. \ref{F:OPsphere}.

The second group of order parameters focuses on microscopic  properties of the underlying CDT triangulations, namely  the shape of the spatial volume profiles $n_t$: 
\beql{E:OP3}
OP_3 = \sum_t \big( n_{t+1} - n_t \big)^2 \ ,
\eeq
and the existence of vertices of very large order: 
\beql{E:OP4}
OP_4 = \max_v O(v) \ ,
\eeq
where $O(v)$ is the vertex coordination number, i.e. the number of simplices sharing a given vertex $v$, thus $OP_4$ is just the coordination number of the highest order vertex in a triangulation. When one looks at the volume profiles $n_t$, see Fig. \ref{F:VolSphere}, one observes that the profile is quite smooth in phase $C_{dS}$ where there are no big differences in $n_t$ and $n_{t+1}$. In phase $C_b$ the volume profile narrows in time direction causing slightly bigger differences between $n_t$ and $n_{t+1}$. The profile is much less smooth in phase $A$, where a kind of "zig-zag" shape  is observed and very non-smooth is phase $B$ where there is a sudden jump in the slice where all spatial volume is concentrated. As a result,  the $OP_3$ is small is phase  $C_{dS}$, medium in phases $A$ and $C_b$, and large in phase $B$, see Fig.~\ref{F:OPsphere}. Last but not least, the existence of high order  vertices is related to the formation of spatial volume clusters which are characteristic for the bifurcation phase $C_b$ \cite{Ambjorn:2015qja,Ambjorn:2017} and  phase $B$ (one huge volume cluster in the collapsed  slice), resulting in large $OP_4$ inside these phases. In phases $C_{dS}$ and $A$ some  spatial volume concentrations may also form from quantum fluctuations but they  do not form any distinguished large-scale structures, making the   $OP_3$  small, see Fig. \ref{F:OPsphere}. The order parameters are summarised in Table \ref{T1}.

\begin{figure}[H]
\centering
\includegraphics[width=0.51\linewidth]{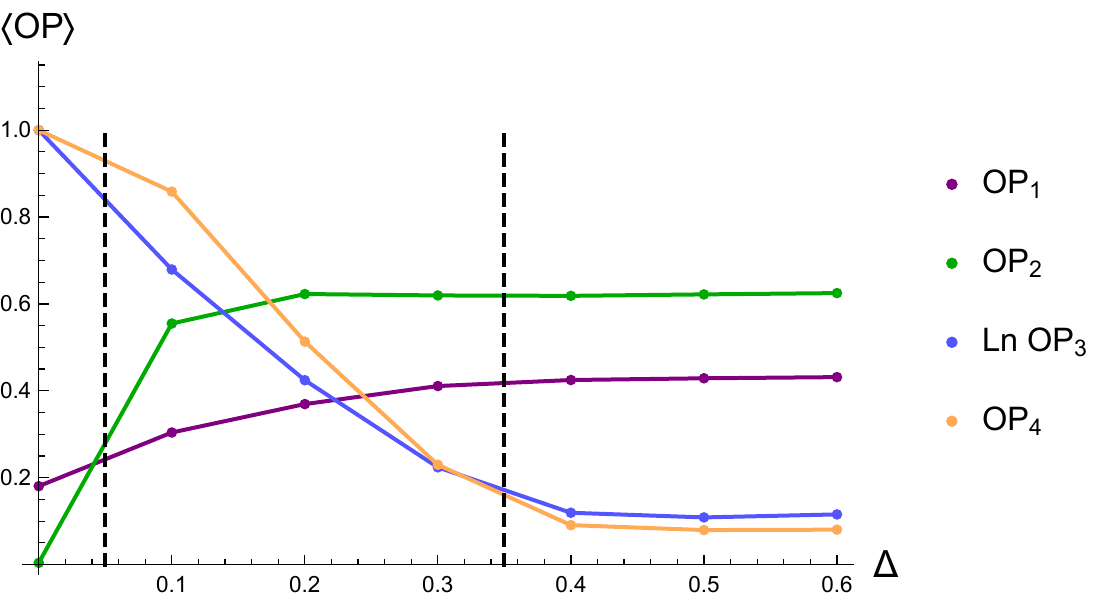}
\includegraphics[width=0.41\linewidth]{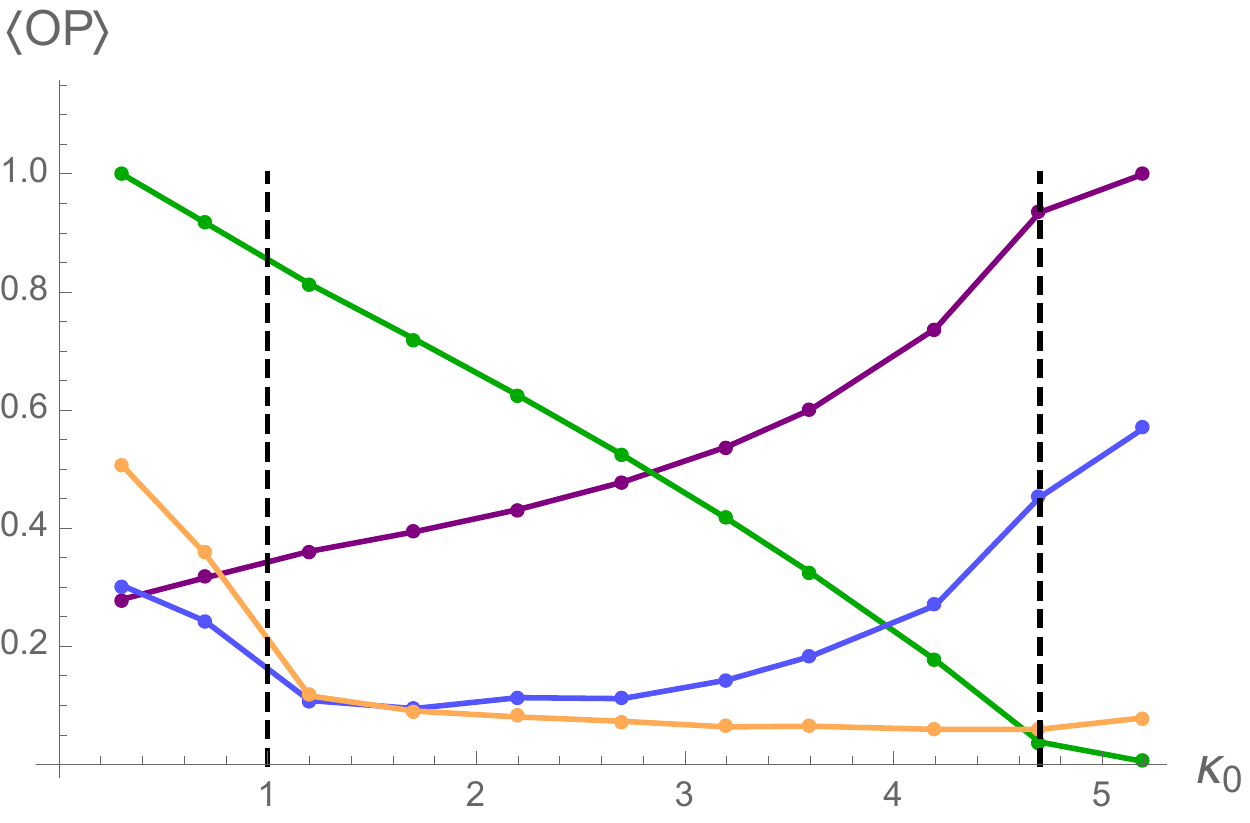}
\caption{\small {Behaviour of the order parameters $OP_1$, ..., $OP_4$, defined in Eqs. (\ref{E:OP1})-(\ref{E:OP4}), in CDT with spherical spatial topology ($\Sigma = S^3$). The left plot shows the OPs as a function of $\Delta$ for fixed $\kappa_0=2.2$, which corresponds to a vertical line in the phase diagram in Fig. \ref{F:phaseold} starting in phase $B$ ($\Delta<0.05$), going through phase $C_b$ ($0.05<\Delta<0.35$), to phase $C_{dS}$ ($\Delta>0.35$). The right plot shows the OPs as a function of $\kappa_0$ for fixed $\Delta=0.6$, which corresponds to a horizontal line in the phase diagram in Fig. \ref{F:phaseold} starting in phase $C_b$ ($\kappa_0< \sim 1.0$), going through phase $C_{dS}$ ($\sim1.0<\kappa_0<4.7$), to phase $A$ ($\kappa_0>4.7$). The $OP_1$, $OP_2$ and $OP_4$ were rescaled to fit into a single plot, and due to a very large range of $OP_3$ (a few orders of magnitude) we plot the rescaled $\ln OP_3$ instead of $OP_3$, the rescaling  in the left plot being identical to that in the right plot.}}
\label{F:OPsphere}
\end{figure}

\begin {table}[H]
\caption {Order parameters used in CDT phase transition studies.} \label{T1} 
\begin{center} 
\begin{tabular} {|c||c|c|c|c|}
\hline
$\mathbf {OP}$ & \bf Phase  $ \mathbf{A}$& \bf Phase  $ \mathbf{B}	$ &\bf Phase  $ \mathbf{C_{dS}}$ &\bf Phase  $ \mathbf{C_{b}}$\\ \hline
\hline
$OP_1$&large&	small &	medium&	medium 	\\ \hline
$OP_2$&small&	small &	large& 	large	 	\\ \hline
$OP_3$&medium&	large &	small&	medium 	\\ \hline
$OP_4$&small&	large &	small& 	large 	\\ \hline
\end{tabular}
\end{center}
\end{table}

We expect that in {the case of}  toroidal spatial topology $\Sigma = T^3$ there exist phases similar to those observed {for}  spherical spatial topology and {that}  we should see a similar behaviour of the order parameters defined above. Thus one can scan  the $(\kappa_0, \Delta)$ parameter space and measure the averages $\langle OP_1\rangle, ...,  \langle OP_4\rangle$ {in order } to identify {the} various phases, if they exist. In order to establish a precise position of the phase transition in the parameter space one can also look at the susceptibilty of an order parameter
\beql{susc}
\chi_{OP} \equiv \langle OP^2 \rangle - \langle OP \rangle^2 \ ,
\eeq 
which should peak at the phase transition point. For each case we will  choose the set of the order parameters which gives the clearest  signal / noise ratio. 
We will also analyze   
\beql{susc1}
\chi_{OP}^{*} \equiv  \frac{\chi_{OP}} {  \langle OP \rangle}
\eeq
or
\beql{susc2}
\chi_{OP}^{**} \equiv \frac{\chi_{OP}} {  \langle OP \rangle^2}.
\eeq
Note that there is always some freedom in choosing the order parameter, and instead of $OP$ one can {use any monotonic function $f(OP)$ of $OP$}. For small fluctuations of  $OP$ around the mean value    
$  \langle OP \rangle$ one has
\beql{suscf}
\chi_{f(OP)} \approx  \big(f'(\langle OP \rangle)\big)^2 \chi_{OP} \ . 
\eeq
Up to a numerical constant, the $\chi_{OP}^{*}$ and $\chi_{OP}^{**}$ are obtained for  $f(OP) = \sqrt{OP}$ and $f(OP) = \ln {OP}$, respectively, and such a choice is useful when an order parameter changes by a few orders of magnitude at a phase transition. The approximation (\ref{suscf}) seems to work very well, see Fig. \ref{F:Ksi} where the results obtained by a redefinition $OP \to f(OP)$ in the raw data and by using approximation (\ref{suscf}) cannot be optically distinguished. 

\begin{figure}[H]
\centering
\includegraphics[width=0.45\linewidth]{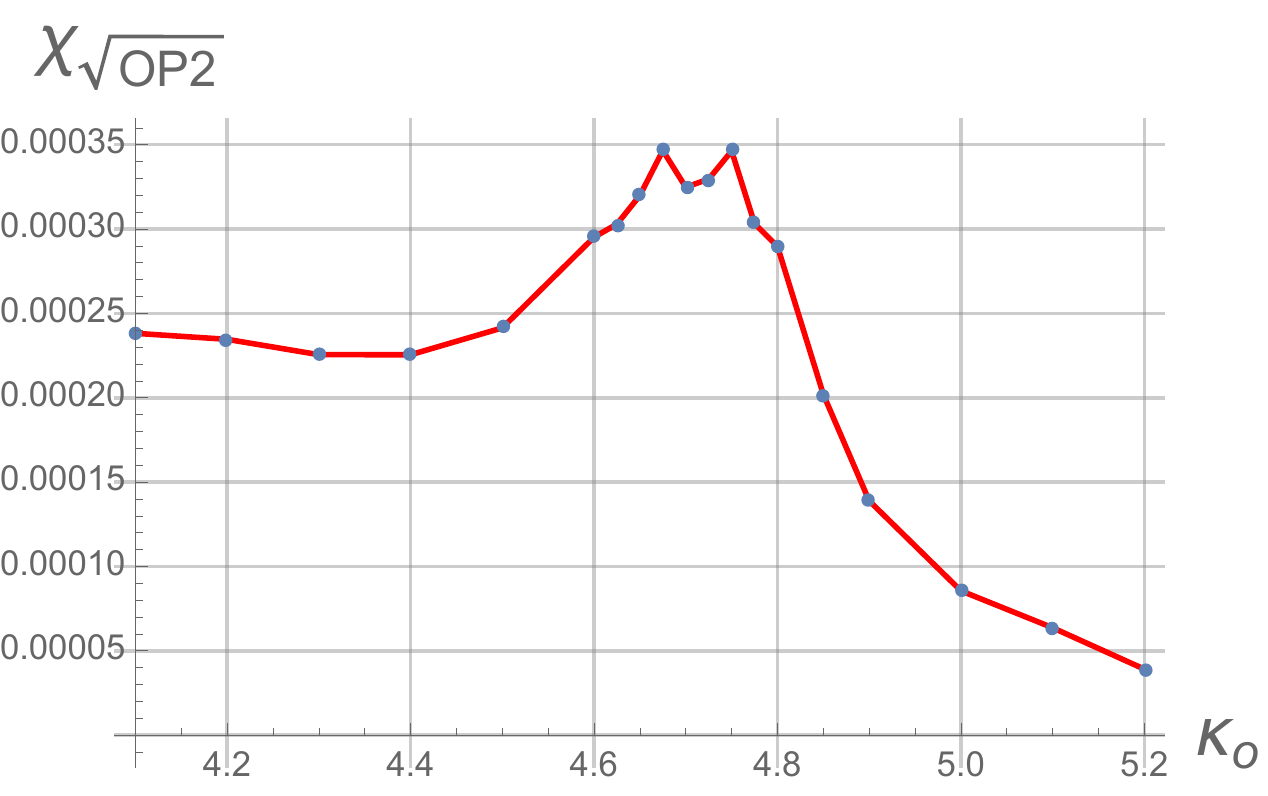}
\includegraphics[width=0.45\linewidth]{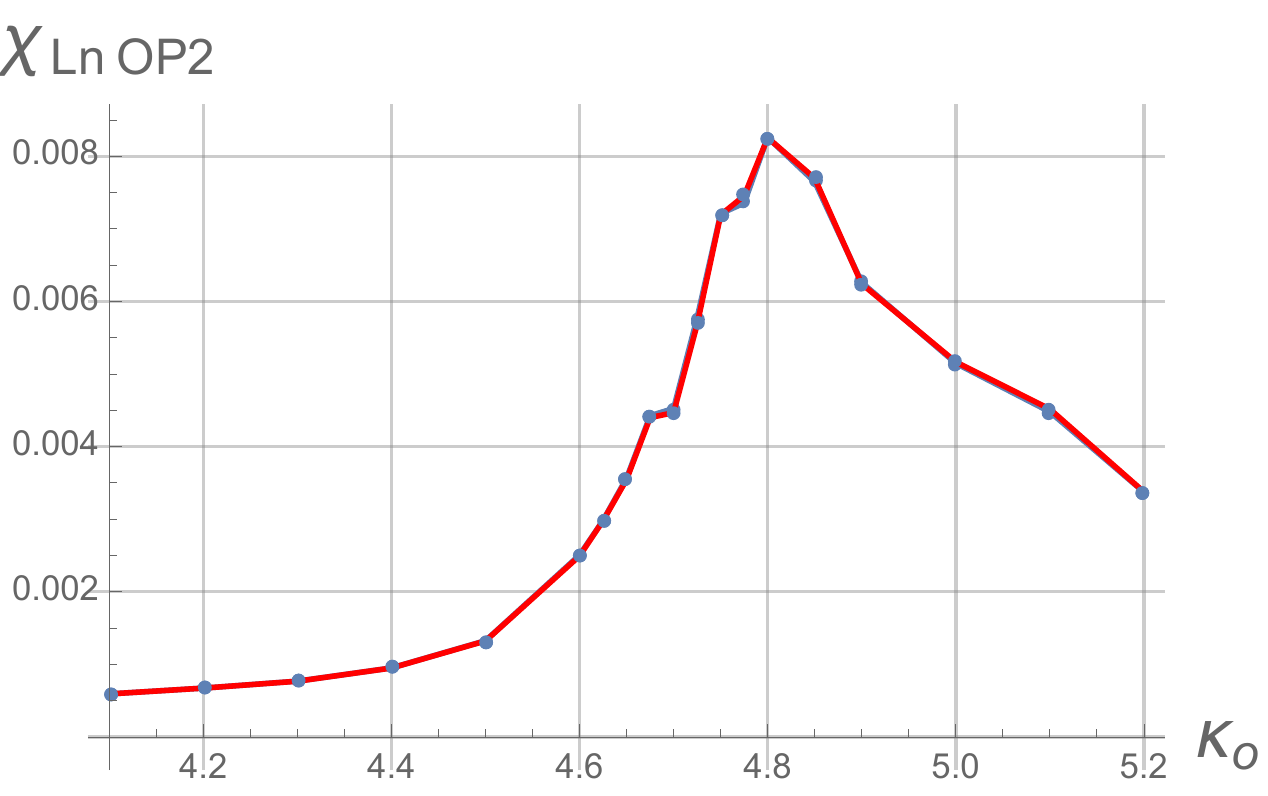}
\caption{\small {Susceptibility of  $\sqrt {OP_2}$ (left chart) and $\ln (OP_2)$ computed directly from raw data (line) and using approximations (\ref{susc1}) and  (\ref{susc2})  (points). Data collected for the $A-C$ phase transition in {the case of}  toroidal topology $\Sigma = T^3$ for fixed $\Delta = 0.3$. }}
\label{F:Ksi}
\end{figure}

\end{section}

\begin{section}{Phase structure in toroidal spatial topology}\label{PhaseStruc}

We begin with a rough scan of the CDT parameter space for fixed spatial topology of a three-torus and time-periodic boundary conditions, i.e. spacetime topology ${\cal M} = T^3 \times S^1$. We {choose} the lattice volume $ \bar N_{(4,1)} = 80000$ and the length of the time period $t_{tot} = 40$. For such a choice of  simulation parameters one   could observe all four CDT phases in the {the case of} spherical topology. We split the parameter space\footnote{The $\kappa_4$ coupling constant is adjusted to the critical value $ \kappa_4^c(\bar N_{(4,1)} ,\kappa_0, \Delta)$, see Section \ref{Numerics} for details.} $(\kappa_0, \Delta)$ into a grid of equally separated points (see Fig. \ref{F:PhasesGrid}) and run simulations in which we measure spatial volume profiles $n_t$ (see Fig. \ref{F:VolTorus}) and the  order parameters $OP_1, ..., OP_4$ described in Section \ref{OrderParam} (see Fig. \ref{F:OPGrid}). 

\begin{figure}[H]
\centering
\includegraphics[width=0.60\linewidth]{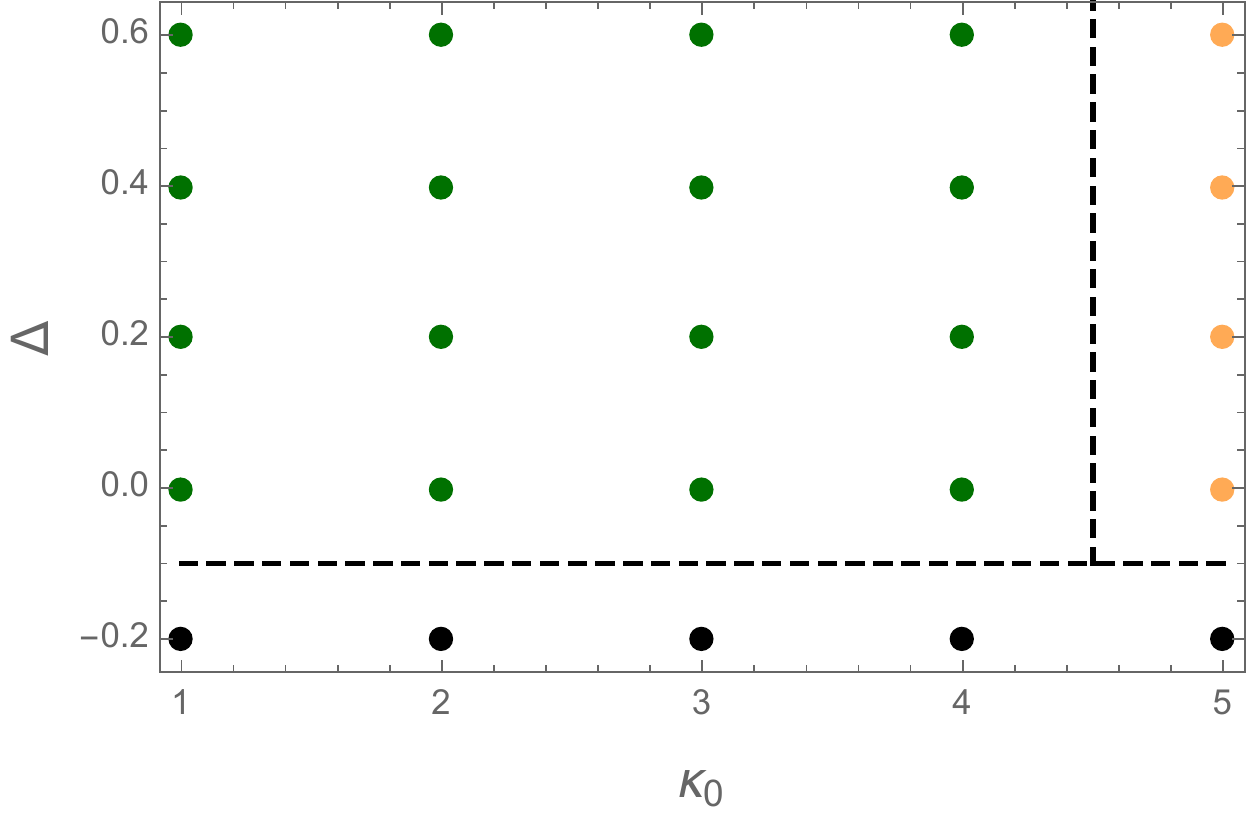}
\caption{\small {Preliminary phase diagram  of the  CDT with toroidal spatial topology ( $\Sigma=T^3$) for $ \bar N_{(4,1)} = 80000$ and $t_{tot} = 40$. Three distinct phases of geometry can be identified. Phase $A$ is observed for large values of $\kappa_0$ (orange points), phase $B$  for small values of $\Delta$ (black  points), and phase $C$ for small $\kappa_0$ and large $\Delta$  (green  points). Approximate position of phase transitions  is denoted by dashed lines.}}
\label{F:PhasesGrid}
\end{figure}

\begin{figure}[H]
\centering
\includegraphics[width=0.3\linewidth]{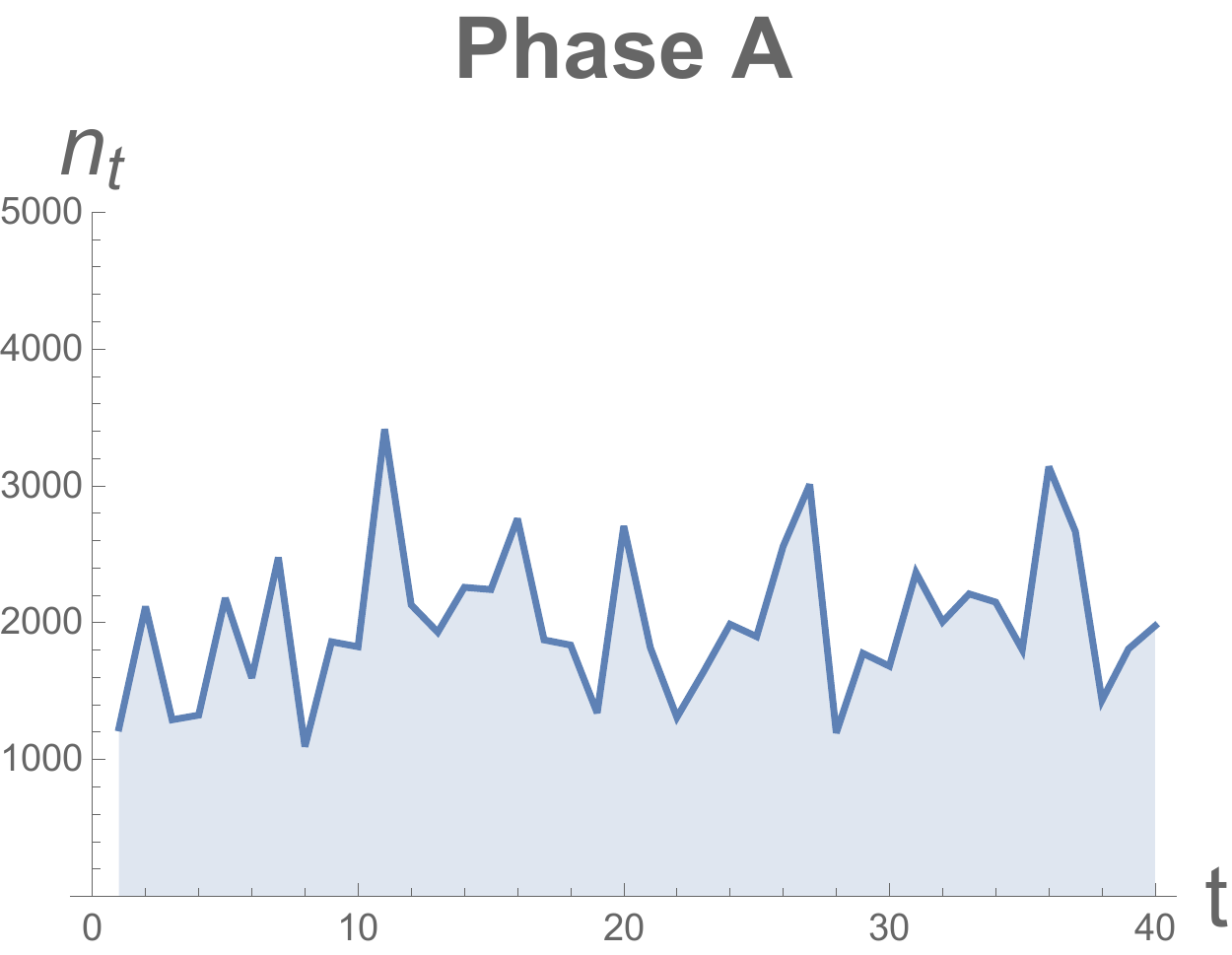}
\includegraphics[width=0.3\linewidth]{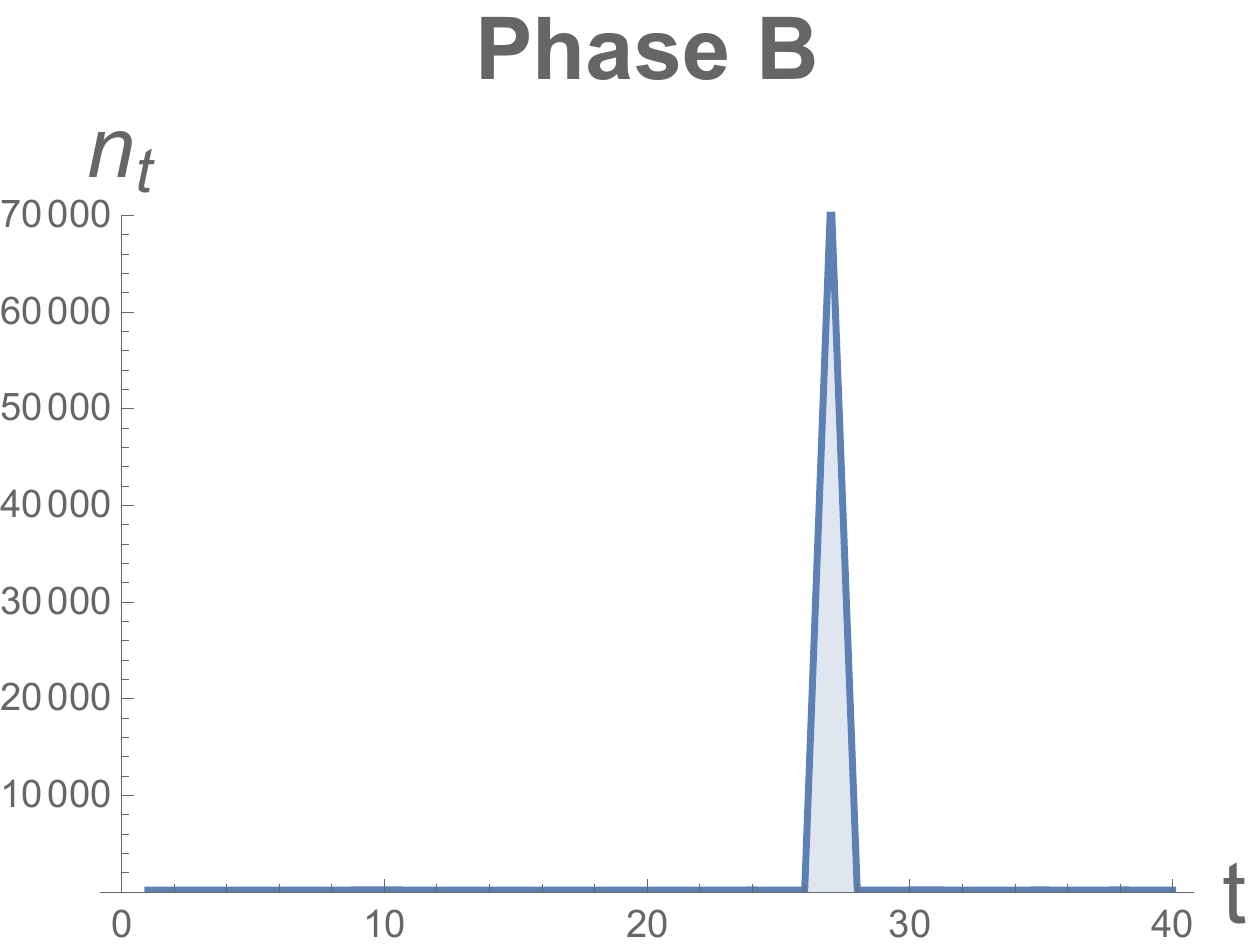}
\includegraphics[width=0.3\linewidth]{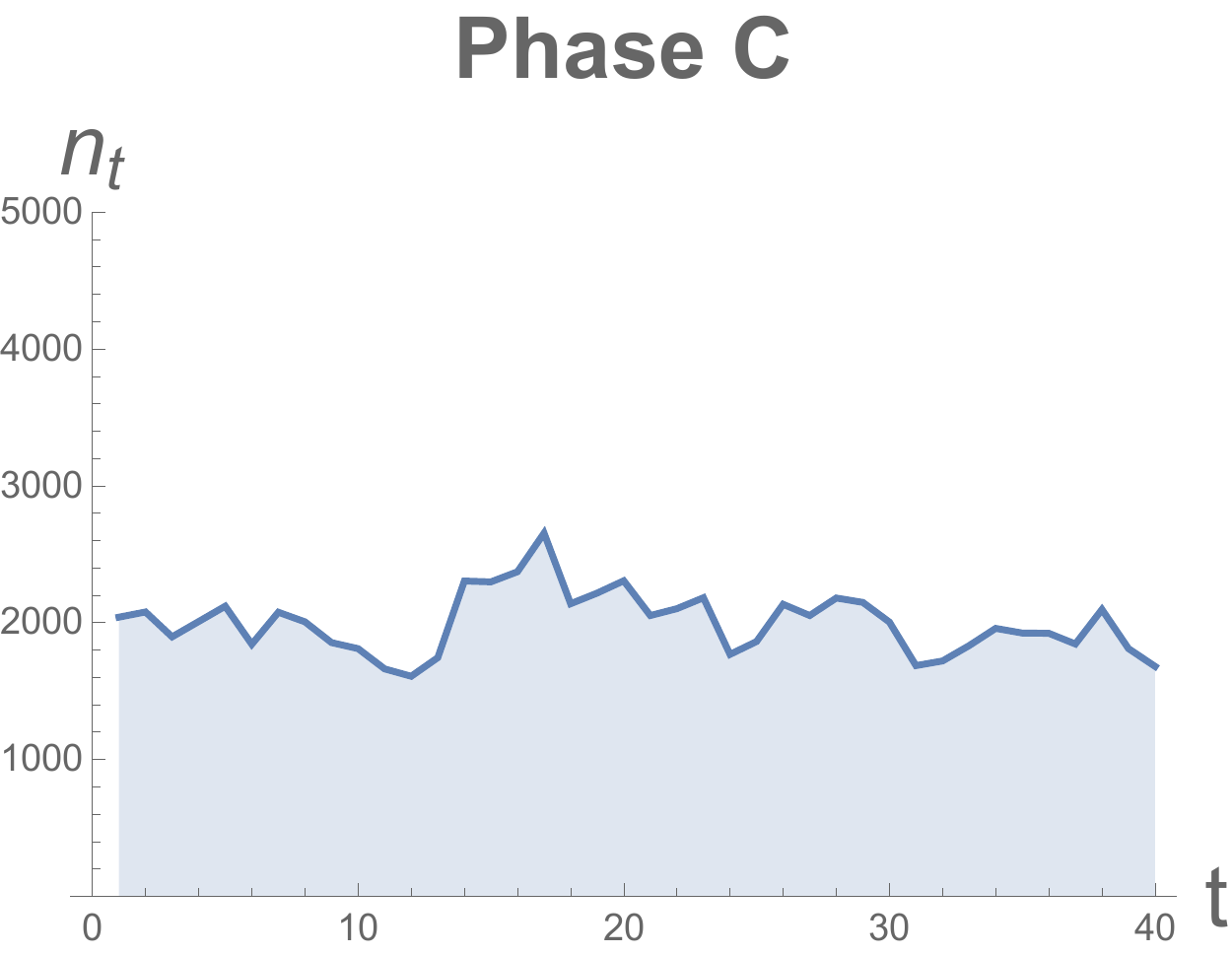}
\caption{\small {Typical volume profiles $n_t \equiv N_{(4,1)}(t)$ in  various CDT phases in toroidal spatial topology $\Sigma=T^3$.  }}
\label{F:VolTorus}
\end{figure}

\begin{figure}[H]
\centering
\includegraphics[width=0.5\linewidth]{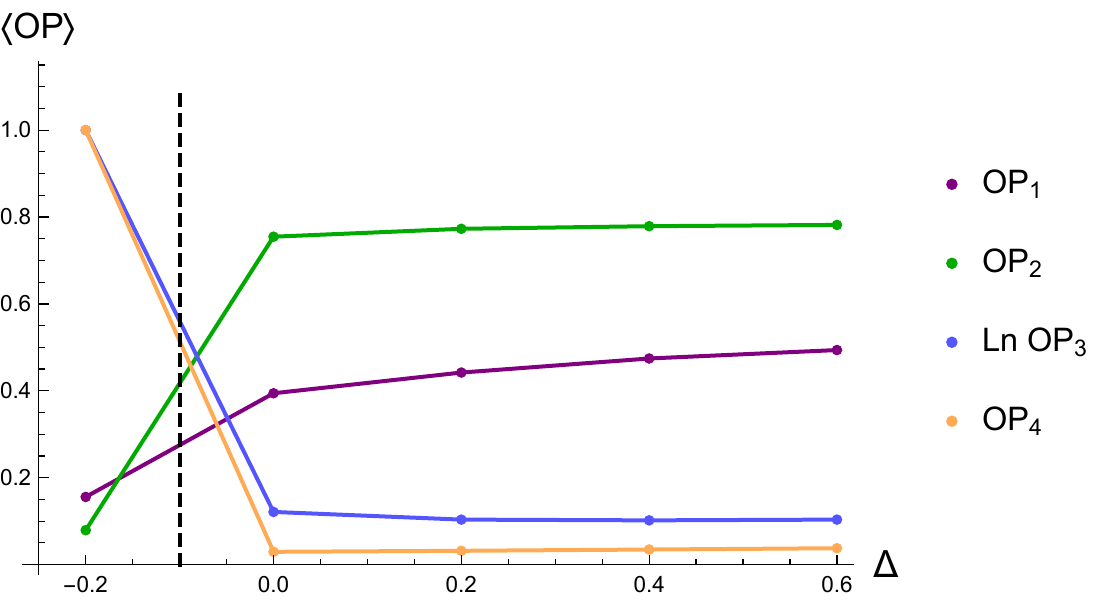}
\includegraphics[width=0.4\linewidth]{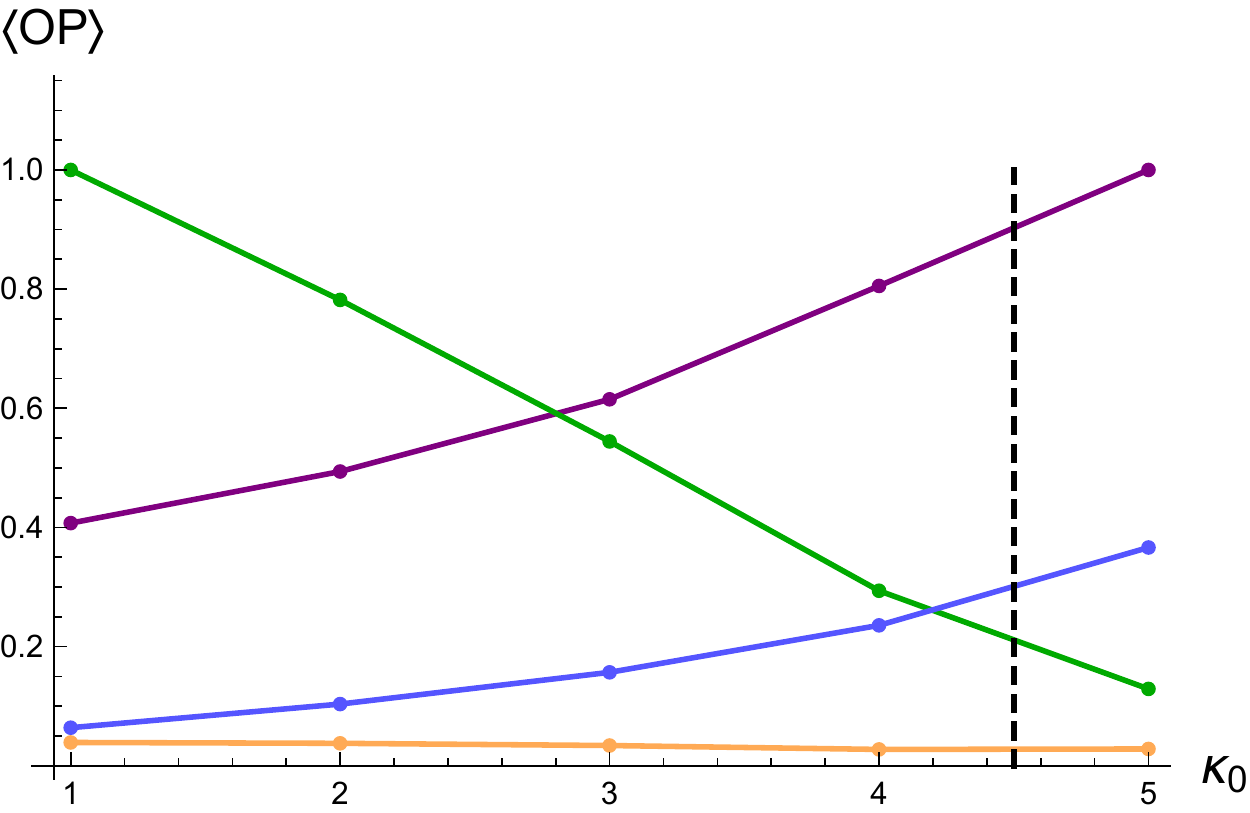}
\caption{\small {Behaviour of the order parameters $OP_1$, ..., $OP_4$, defined in Eqs. (\ref{E:OP1})-(\ref{E:OP4}), in CDT with toroidal spatial topology ($\Sigma = T^3$) and $ \bar N_{(4,1)} = 80000$, $t_{tot} = 40$. {The} left plot shows the OPs as a function of $\Delta$ for fixed $\kappa_0=2.0$, which corresponds to a vertical line in the phase diagram in Fig. \ref{F:PhasesGrid}, and the  right plot shows the OPs as a function of $\kappa_0$ for fixed $\Delta=0.6$, which corresponds to a horizontal line in the phase diagram in Fig. \ref{F:PhasesGrid}. The $OP_1$, $OP_2$ and $OP_4$ were rescaled to fit into a single plot, and due to a very large range of $OP_3$ (a few orders of magnitude) we plot the rescaled $\ln OP_3$ instead of $OP_3$, the rescaling  in the left  and { right plots being the same}. {The} qualitative behaviour of the OPs is the same as for phases $A$, $B$ and $C_{dS}$ in CDT with spherical spatial topology (see Fig. \ref{F:OPsphere} and Table \ref{T1}). Approximate positions of the $B-C$ and the  $C-A$ phase transitions are marked by dashed lines on the left and right plots, respectively.}}
\label{F:OPGrid}
\end{figure}

By looking at these data one can easily notice three distinct phases of spacetime geometry, denoted $A$, $B$ and $C$. The position of phases $A$ and $B$ on the  phase diagram in Fig. \ref{F:PhasesGrid}, as well as  the spatial volume profiles (series of uncorrelated spatial slices in phase $A$, and time-collapsed volume structure in phase $B$) and values of the order parameters (see Fig. \ref{F:OPGrid} and Table  \ref{T1}) suggest that the phases are in one-to-one correspondence to phases $A$ and $B$  of CDT with spatial topology $\Sigma=S^3$ (see Fig. \ref{F:phaseold}). 

Existence of phase $C$ in  spatial topology $\Sigma=T^3$ was already reported in \cite{Ambjorn:2016fbd,Ambjorn:2017226}, where the authors noticed that the spatial volume profile $n_t$ is highly correlated and can be well described by a toroidal minisuperspace-like action with small quantum corrections. It can be argued that this is the toroidal analogue of the semi-classical  phase $C_{dS}$, earlier observed for the  spatial topology $\Sigma=S^3$ (see Fig. \ref{F:phaseold}). This is further confirmed by  behaviour of the order parameters  $OP_1, ..., OP_4$ (see Fig. \ref{F:OPGrid} and Table \ref{T1}).

The fact that we don't see the toroidal analogue of the bifurcation phase $C_b$ in the above data is not very surprising as the (expected) constant volume profile of spatial slices  causes the volume of a single slice to be too small ($n_t \sim \bar N_{(4,1)} / t_{tot} = 2000$) to allow for a creation of large volume clusters and consequently for the emergence of  high-order vertices. The same effect appeared in the spherical topology for small systems where high order vertices could be observed only in  slices with spatial volume higher than the, so-called bifurcation point volume. In CDT with spherical spatial topology the volume profile $n_t$ had a characteristic blob structure (see Fig. \ref{F:VolSphere}) with the central part volume much higher than the average volume $\bar N_{(4,1)} / t_{tot}$ and thus the choice of  $\bar N_{(4,1)} =80000$ and $t_{tot} = 40$ was good enough to let the central part volume be higher than the bifurcation point volume. As a result the high-order vertices could form inside the blob part of the spherical CDT triangulations, which seems not to be the case in the toroidal CDT  with flat volume profiles.

To circumvent  this obstacles, i.e. to let the average spatial volume of the toroidal CDT triangulations exceed the bifurcation point volume, 
we decided to pursue a detailed study of the toroidal CDT phase structure for much bigger average volume $\bar N_{(4,1)} / t_{tot} = 40000$, by setting  $\bar N_{(4,1)}= 160000$ and $t_{tot} = 4$, respectively. To use our computer resources effectively, we have also carefully fine-tuned the grid of the $(\kappa_0, \Delta)$  points in which we performed  numerical simulations, such that we have much higher precision near expected phase transition points, see Fig. \ref{F:phases} where the phase diagram is shown. 
\begin{figure}[h]
\centering
\includegraphics[width=0.70\linewidth]{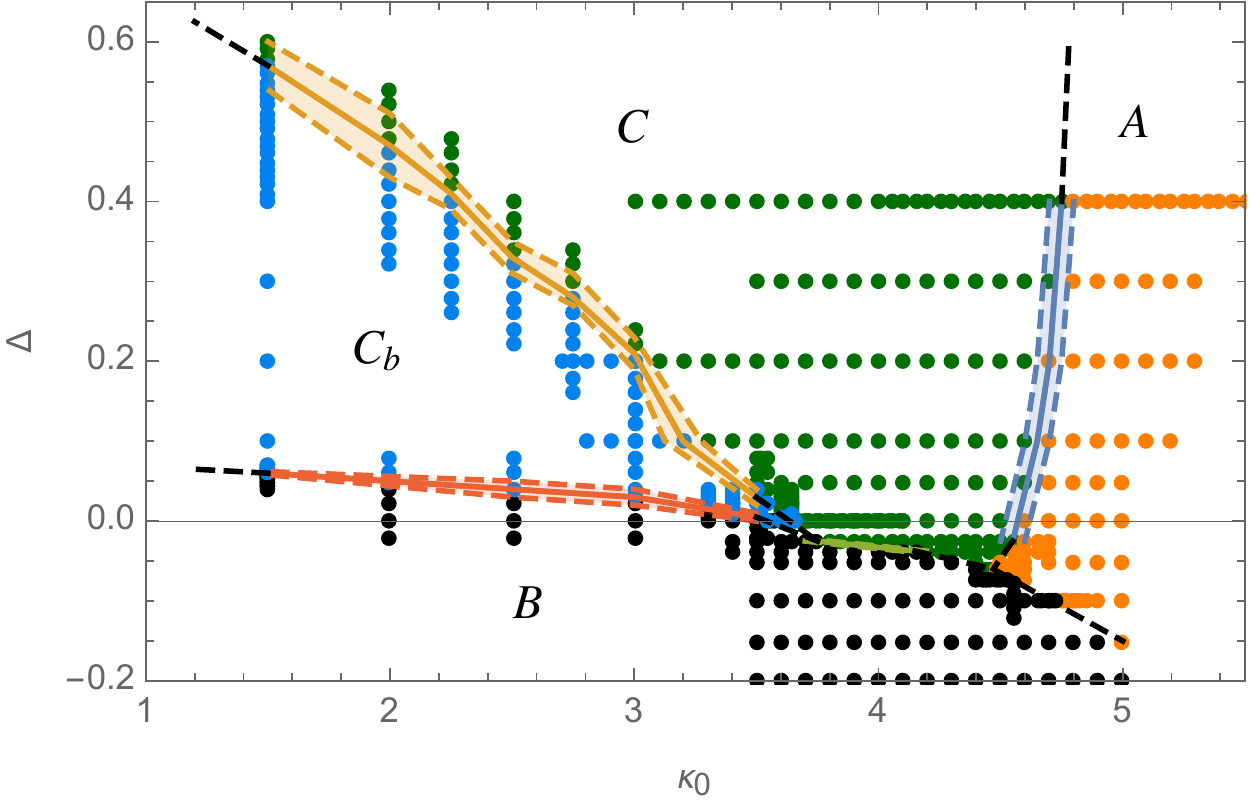}
\caption{\small {Phase diagram of  CDT with toroidal spatial topology ( $\Sigma=T^3$) for $ \bar N_{(4,1)} = 160000$ and $t_{tot} = 4$.  Points denote the actual  numerical simulations. {The} measured phase transitions are marked by  solid lines with error bars denoted by shaded areas of the same color. The error bars for the $C-C_b$ transition are due to the observed hysteresis of the measured order parameters (see Section \ref{Discussion} for discussion), the error bars for {the} other phase transitions are purely due to the resolution of the measurement grid. Dashed black lines are extrapolations of the measured phase transition lines. All four  distinct phases of geometry can be identified. Phase $A$ is observed for large values of $\kappa_0$ (orange points), phase $B$  for small values of $\Delta$ (black  points),  phase $C_b$ for small $\kappa_0$ and medium $\Delta$  (blue points), and phase $C$ for small $\kappa_0$ and large $\Delta$  (green  points).}}
\label{F:phases}
\end{figure}
These results reconfirm  {the} findings of  the initial phase transition study from Fig. \ref{F:PhasesGrid}, and  also confirm the existence of the fourth phase, denoted  $C_b$ (blue points in Fig. \ref{F:phases}).
 The phase is again a toroidal analogue of the bifurcation phase $C_b$ observed in the spherical topology, with  spatial homogeneity broken by a formation of  volume clusters around high order vertices
 emerging in the every second spatial layer, see Fig.~\ref{F:Omax}.
The behaviour of all four order parameters as functions of $\kappa_0$ and $\Delta$ is shown in Fig. \ref{F:OPphases}  and their susceptibilities in Fig. \ref{F:Chiphases}. These results  can be used to  draw the phase transition lines with high precision, as presented in Fig.  \ref{F:phases}. The phase structure looks very similar to the one observed for a spherical spatial topology, see Fig. \ref{F:phaseold}.

 \begin{figure}[H]
\centering
\includegraphics[width=0.55\linewidth]{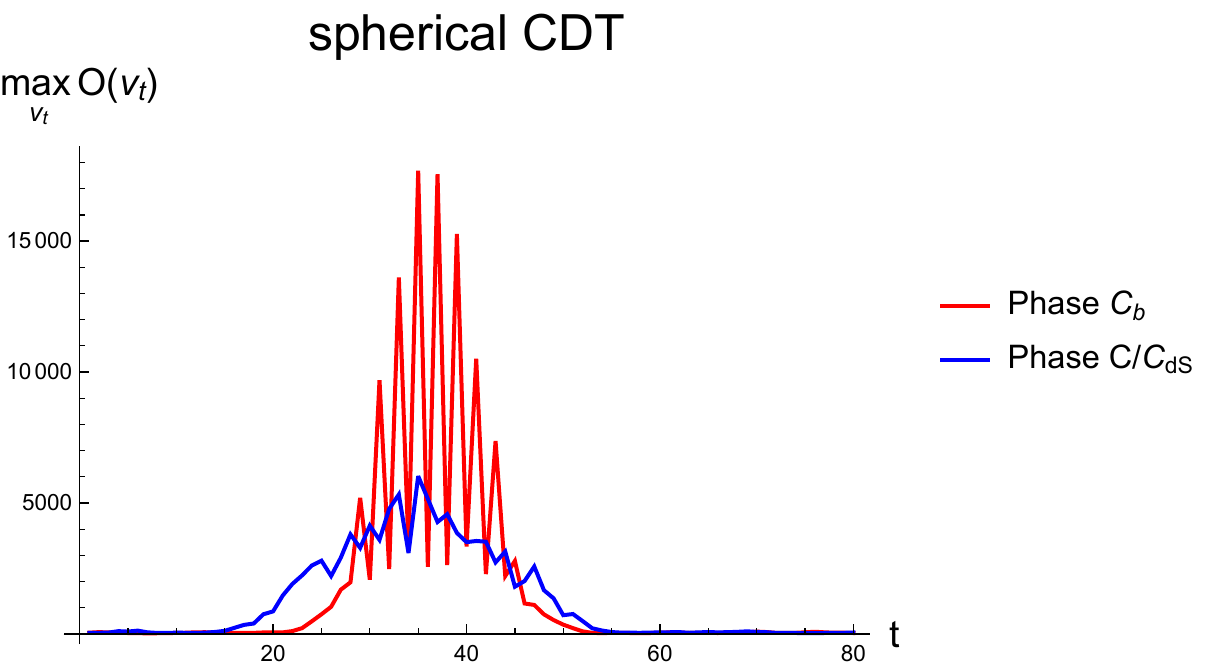}
\includegraphics[width=0.4\linewidth]{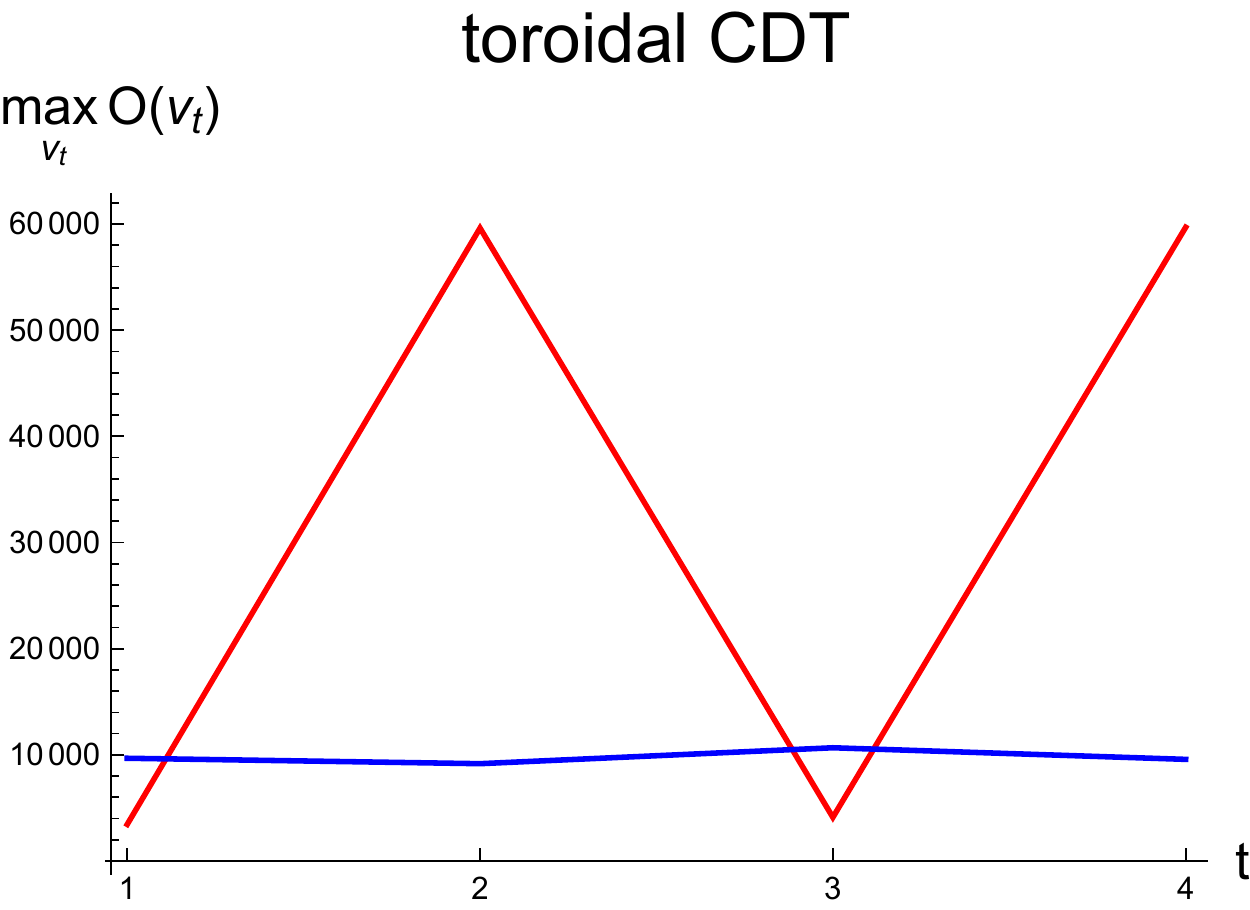}
\caption{\small {Time structure of highest order vertices in  CDT triangulations with spherical (left plot) and toroidal (right plot) spatial topology. $\max_{v_t} O(v_t)$ denotes the coordination number of the highest order vertex with time coordinate $t$. Distinction between the bifurcation phase $C_b$ with characteristic high order vertices in every second time layer, and phase $C/C_{dS}$ is clear in both topologies. }}
\label{F:Omax}
\end{figure}

\begin{figure}[H]
\centering
\includegraphics[width=0.5\linewidth]{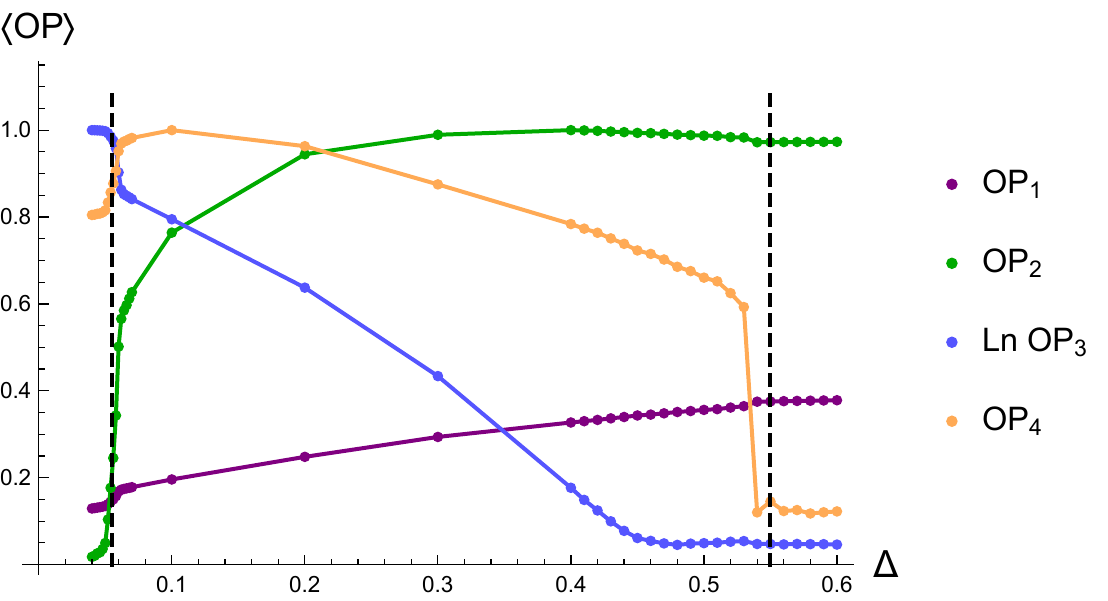}
\includegraphics[width=0.4\linewidth]{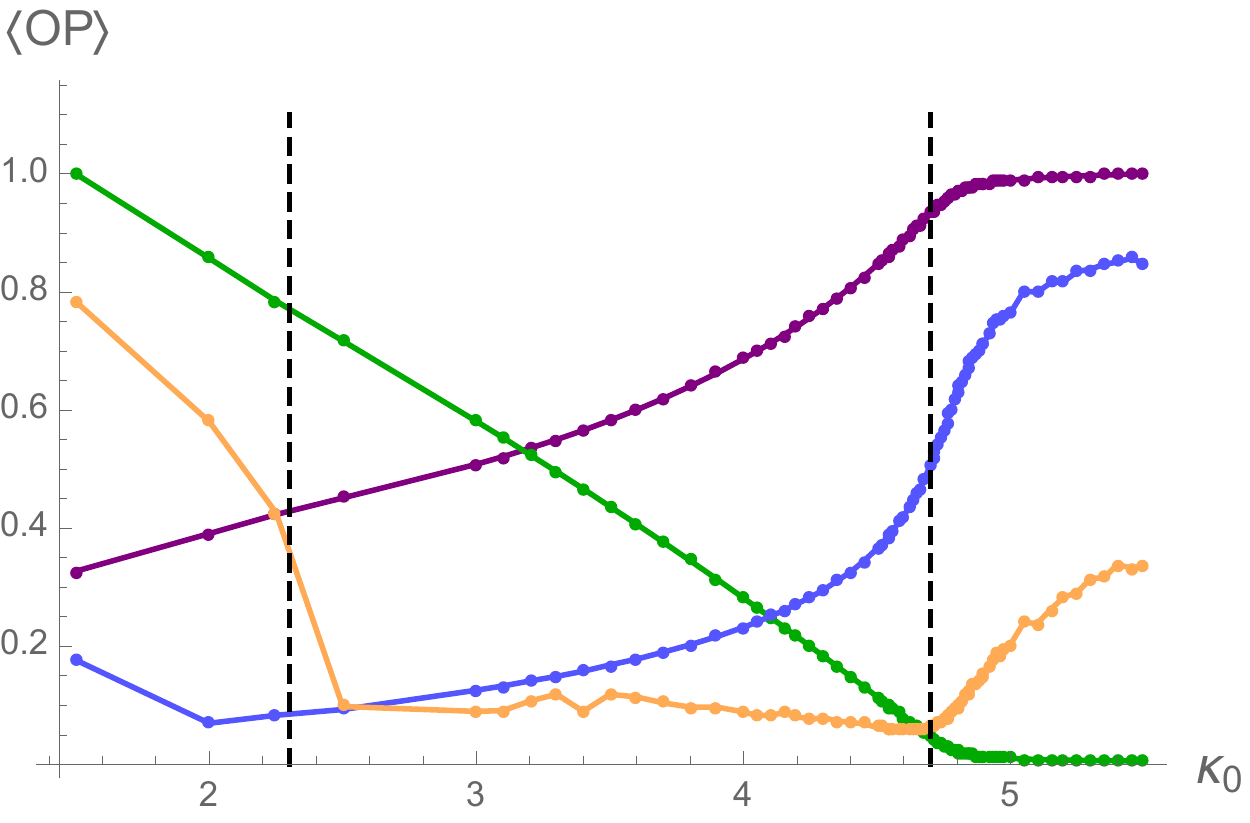}
\caption{\small {Behaviour of the order parameters $OP_1$, ..., $OP_4$, defined in Eqs. (\ref{E:OP1})-(\ref{E:OP4}), in CDT with toroidal spatial topology ($\Sigma = T^3$) and $ \bar N_{(4,1)} = 160000$, $t_{tot} = 4$. {The} left plot shows the mean OPs as a function of $\Delta$ for fixed $\kappa_0= 1.5 $, which corresponds to a vertical line in the phase diagram in Fig. \ref{F:phases}, and the  right plot shows the mean OPs as a function of $\kappa_0$ for fixed $\Delta=0.4$, which corresponds to a horizontal line in the phase diagram in Fig. \ref{F:phases}. The $OP_1$, $OP_2$ and $OP_4$ were rescaled to fit into a single plot, and due to a very large range of $OP_3$ (a few orders of magnitude) we plot the rescaled $\ln OP_3$ instead of $OP_3$, the rescaling  in the left plot {being} identical {to that} in the right plot. {The} qualitative behaviour of the OPs is the same as for phases $A$, $B$, $C_{dS}$ and $C_b$ in CDT with spherical spatial topology (see Fig. \ref{F:OPsphere} and Table \ref{T1}).  {The} positions of the $B-C_b$ and $C_b-C$  phase transitions on the left plot and  the  $C_b-C$ and $C-A$ phase transition  on the right plot are marked by dashed lines.}}
\label{F:OPphases}
\end{figure}

\begin{figure}[H]
\centering
\includegraphics[width=0.5\linewidth]{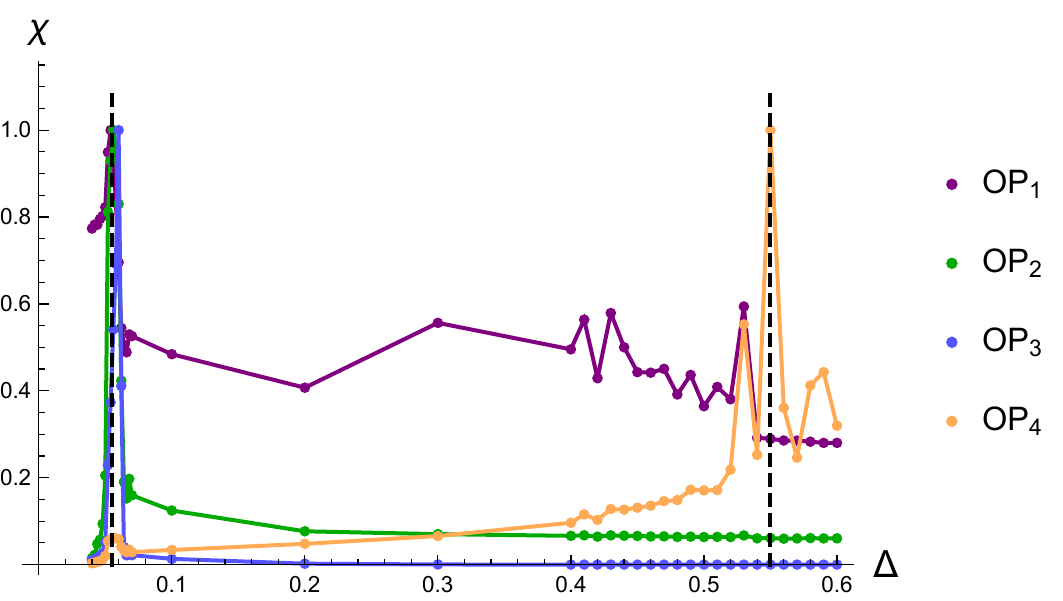}
\includegraphics[width=0.4\linewidth]{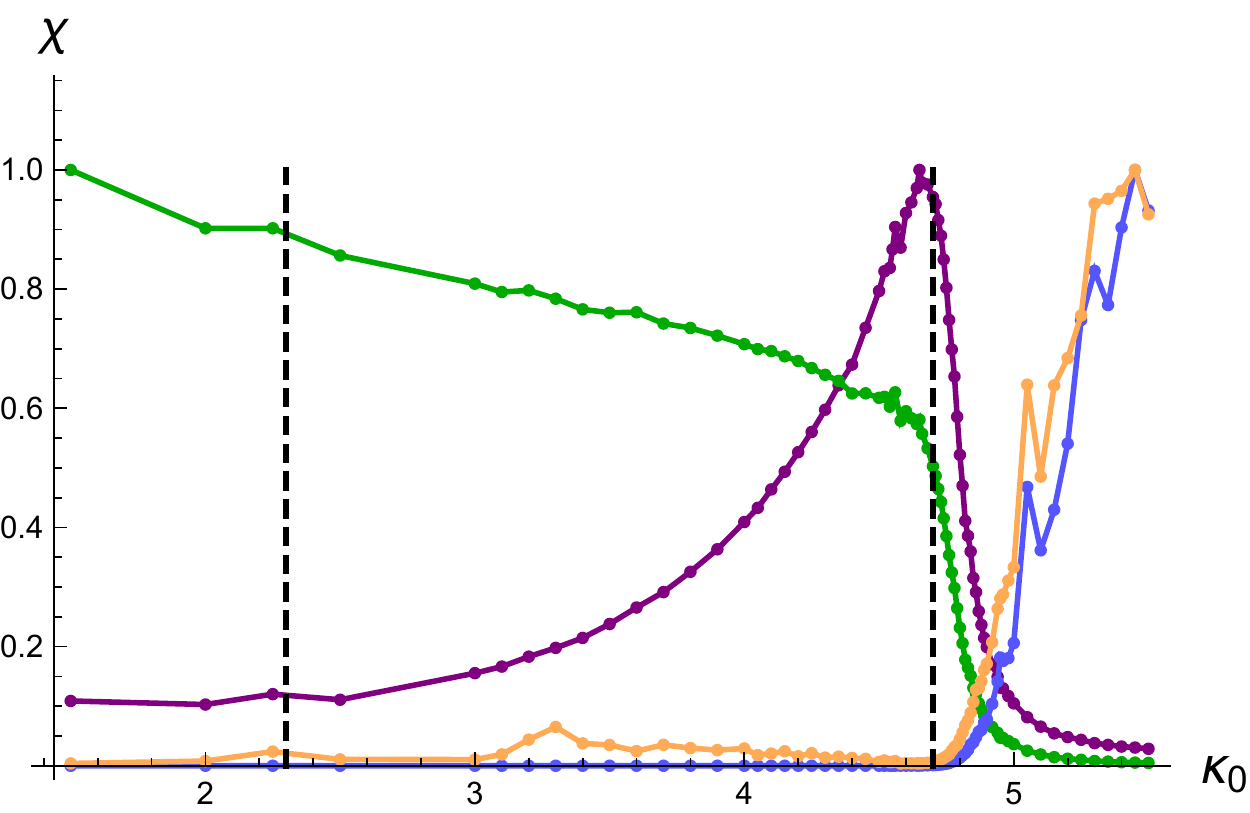}
\includegraphics[width=0.5\linewidth]{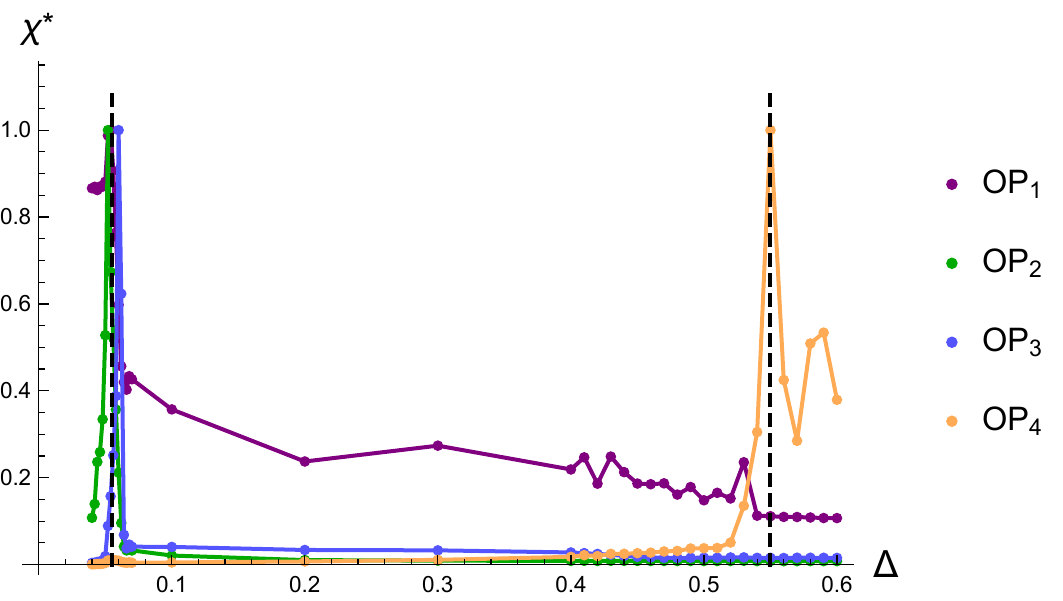}
\includegraphics[width=0.4\linewidth]{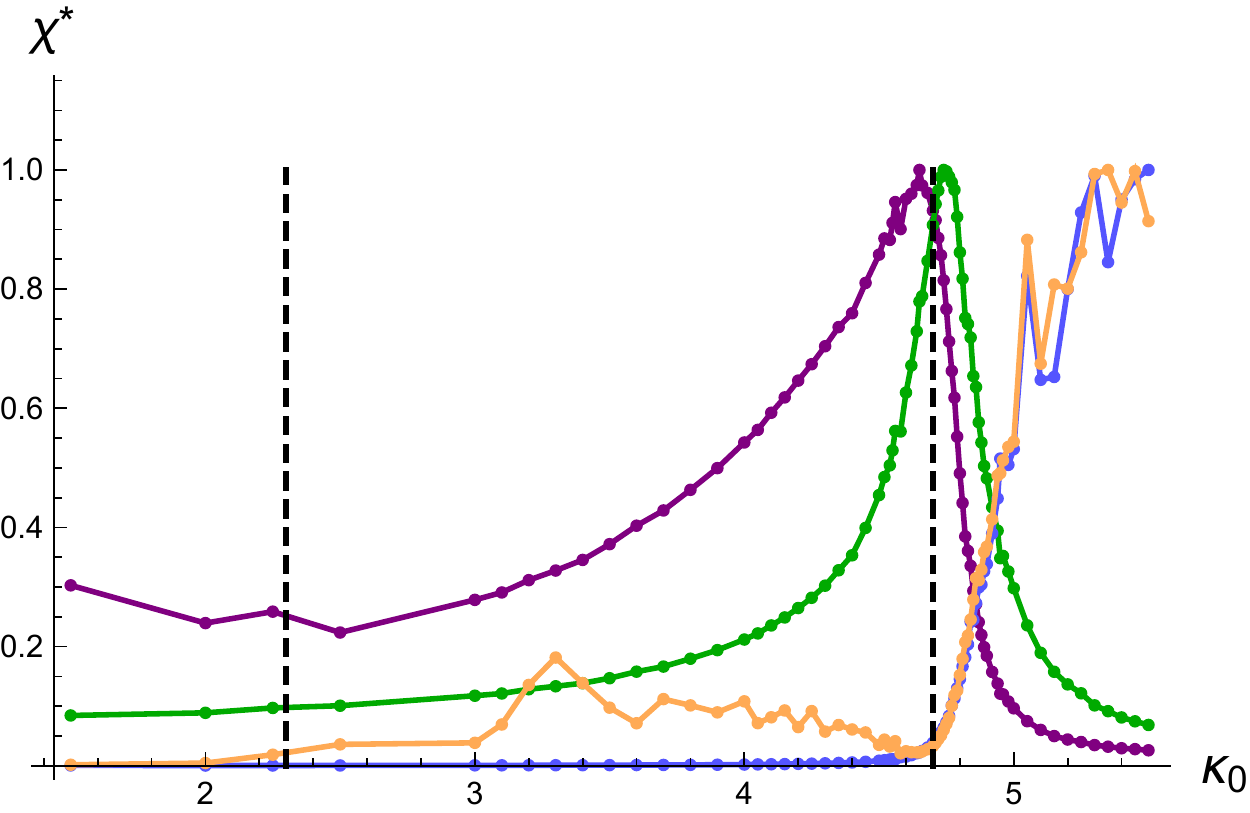}
\caption{\small {Susceptibilities $\chi$  (top plots) and $\chi^*$ (bottom plots) of the order parameters $OP_1$,~..., $OP_4$, defined in Eqs.~(\ref{susc}) and (\ref{susc1}), respectively, in CDT with toroidal spatial topology ($\Sigma = T^3$) and $ \bar N_{(4,1)} = 160000$, $t_{tot} = 4$. {The} left plots show the susceptibilities as a function of $\Delta$ for fixed $\kappa_0= 1.5 $ and the  right plot shows the susceptibilities as a function of $\kappa_0$ for fixed $\Delta=0.4$, which corresponds to the order parameters shown in Fig. \ref{F:OPphases}. The susceptibilities were rescaled to fit into a single plot. {The} positions of the $B-C_b$ and $C_b-C$  phase transitions on the left plot and  the  $C_b-C$ and $C-A$ phase transition  on the right plot are signaled by peaks in {the} susceptibilities, which is marked by the same dashed lines as in Fig. \ref{F:OPphases}.}}
\label{F:Chiphases}
\end{figure}

\end{section}

\begin{section}{Discussion and conclusions}\label{Discussion}
 
 \begin{figure}[h]
\centering
\includegraphics[width=0.70\linewidth]{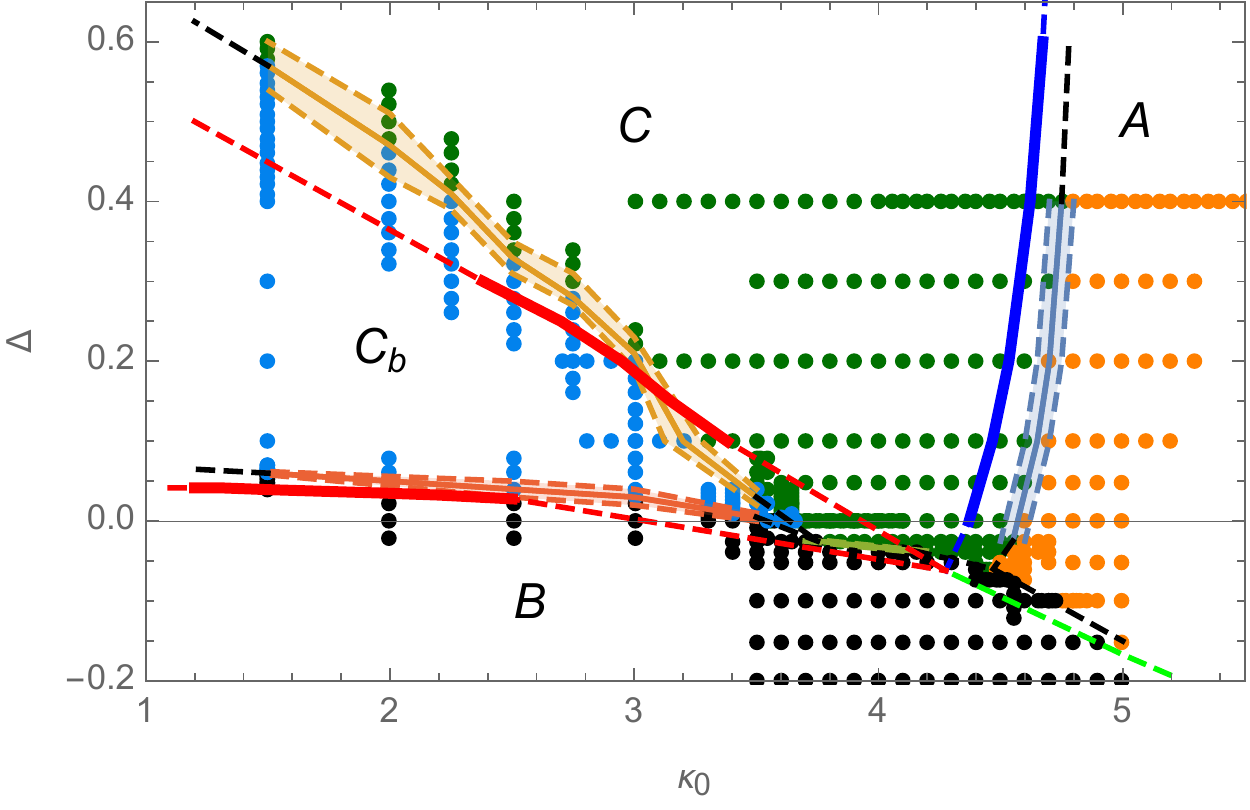}
\caption{\small {{Phase diagram of  CDT with toroidal spatial topology from Fig.~\ref{F:phases} together with the phase transition lines (thick lines) of CDT with spherical spatial topology from Fig.~\ref{F:phaseold}. }}}
\label{F:phasescomb}
\end{figure}
 
{We  studied the phase diagram of  four-dimensional CDT with toroidal spatial topology ($\Sigma=T^3$) and time periodic boundary conditions, see Fig. \ref{F:phasescomb}. Our results  confirm the existence of four distinct phases of quantum geometry which are direct analogues of the phases previously observed for the spherical spatial topology ($\Sigma= S^3$). The position of the critical lines 
 in the $(\kappa_0, \Delta)$ parameter space  is also very similar in both topologies. The $A-C$ and $B-C_b$ phase transition  lines measured in the toroidal case are slightly shifted compared to the lines measured in the spherical case. Using our present data one cannot completely exclude that the shifts are  real effects which might be attributed to the topology change. However it is much more likely that the shifts are due to final size effects as the positions of (pseudo-)critical points in the parameter space depend on the lattice volume  and  similar parallel shifts were observed in the spherical case when the lattice volume was increased \cite{Ambjorn:2011cg,Ambjorn:2012ij,Coumbe:2015oaa,Ambjorn:2017nho}. On the one hand, the finite size effects should be much stronger in the toroidal CDT, where the minimal triangulation is much larger than in the case of the spherical CDT \cite{Ambjorn:2016fbd}. On the other hand, the length of the (periodic) time axis used in the toroidal CDT simulations presented herein was much shorter than in the spherical simulations, and the resulting effective volume per slice was much larger in the toroidal case.}

{The critical region near the point where the phase transition lines meet is the most interesting place to concentrate on, since, following the asymptotic safety arguments, it is a natural candidate for the physical UV limit of CDT  \cite{Ambjorn:2014gsa,Ambjorn:2016cpa}. This region could not be studied with a sufficiently high precision for the spherical topology. The reason was purely
technical: the local Monte Carlo algorithm we use to update triangulations was  very inefficient in this critical region. As a result we couldn't make  precise measurements there and our conjecture about the existence of a common "quadruple point" where all four phases meet was based on an extrapolation of the measured phase transition lines.   The situation is different for the
toroidal case, where the Monte Carlo algorithm works fine in the critical "corner" region of the parameter space, as can be seen from a plot shown in Fig.  \ref{F:phasescomb}. In this case the common "quadruple point" seems less likely.
The $C_b-C$ phase transition line  is now shifted slightly to the left and tilted in the $(\kappa_0, \Delta)$ plane 
compared to the $C_b-C_{ds}$ phase transition  line observed in the case of spherical spatial topology. Consequently, in the toroidal case, it seems more likely that we have the "old" triple point around  $(\kappa_0=4.50, \Delta=-0.05)$ where $A$, $B$ and $C$ phases meet and a "new" triple point around $(\kappa_0=3.75, \Delta=-0.02)$ where the bifurcation phase
$C_b$  meets the phases $B$ and $C$.  As a result there exists a region in the parameter space, where one can observe a direct transition line between the semiclassical  phase  $C$ and the collapsed phase $B$, see Fig. \ref{F:phasescomb}. We will concentrate our future precise measurements in the "critical corner" region 
to determine accurately the phase diagram. Although the grid of measured points presented in the article
seems rather dense, we cannot exclude that the exact shape of transition lines is more complicated
in the sense that they may bend and meet in one "quadruple point" (again see Fig. \ref{F:phasescomb}). In this precise study we will also measure a sequence of volumes which should enable us to analyze finite size effects affecting the infinite volume position of the phase transition lines more accurately. The results will be published in forthcoming publications.}

In the data presented above we did not measure the order of the phase transitions. Such a study would require massive numerical simulations to be performed for a suitable choice of various lattice sizes enabling one to extrapolate the results to the infinite  volume limit and to measure critical exponents. 
{Nevertheless we  have made some initial observations.}

{First, in  CDT with toroidal spatial topology  one can observe a clear $C-A$ transition signal (peak in susceptibilities) both for the $OP_1$ and $OP_2$  parameters, as it was the case for the spherical CDT. Nevertheless the behaviour of the parameters at the phase transition is quite different from the behavior originally observed  for CDT with spherical spatial topology
where one could observe that the order parameters jump  between two clearly separated metastable states, one for phase $A$ and one for phase $C_{dS}$ \cite{Ambjorn:2012ij}. This suggested it was the first-order transition 
and this was confirmed by a detailed finite size analysis. In the toroidal case the order parameters change smoothly between the two phases and one does not observe any separation of  states. While this could be an indication of 
a higher order transition, it is more likely that it reflects that one is using a constraint of the four-volume (namely 
\rf{Vfix} where $N_{(4,1)}$ is kept fixed) different  than that used in \cite{Ambjorn:2012ij} (where $N_4$ was kept fixed). A similar phenomena were observed 
for the $B-C_b$ transition \cite{Ambjorn:2017}. Clearly the only way to settle the issue is to perform 
a carefull finite size analysis to determine the order of the transition.}

{Secondly, the $C_b-C_{dS}$ phase transition in case of CDT with spherical spatial topology was found to be a second (or higher) order transition \cite{Coumbe:2015oaa,Ambjorn:2017nho}. Now, in the toroidal case, one can observe a clear hysteresis of all  measured order parameters when moving from phase $C$ to phase $C_b$ or the opposite.  There is a large region in the parameter space, denoted by the orange shaded area between phases $C$ and $C_b$ in Fig.~\ref{F:phasescomb}, where the value of an order parameter depends on its initial value, or more precisely on the  geometric configuration (triangulation) used to initiate Monte Carlo simulations. If one starts with an initial triangulation from phase $C$ and makes simulations inside the hysteresis region the generated  triangulations persist in phase $C$, if one starts instead with a triangulation from phase $C_b$ the system persists in phase $C_b$, even for very long simulation runs (a few months of CPU time, or a few $ \times 10^{12}$ attempted Monte Carlo moves). Of course if one goes deep enough into phase $C$ or alternatively into phase $C_b$ (outside the shaded region in Fig. \ref{F:phases}) the system finally thermalizes to phase $C$ or phase $C_b$, respectively, independent on the starting configuration. This is illustrated in Fig.~\ref{F:Hysteresis}, where we show the OPs as a function of $\Delta$ for fixed $\kappa_0= 2.0$. For each $\Delta$ we plot the measurements done in many independent Monte Carlo runs with different initial configurations. The hysteresis region  between phases $C$ and $C_b$ is clearly visible for $\Delta\geq0.38$. Again, while this can indicate a first order transition, and thus a change of 
transition order with topology, it could also be an algorithmic issue with the Monte Carlo simulations since our configurations
are relatively small and the toroidal topology clearly is more constraining than the spherical topology. Again the only 
way to settle the issue is to perform a proper finite size analysis, which is of course quite computer  demanding.} 

\begin{figure}[H]
\centering
\includegraphics[width=0.7\linewidth]{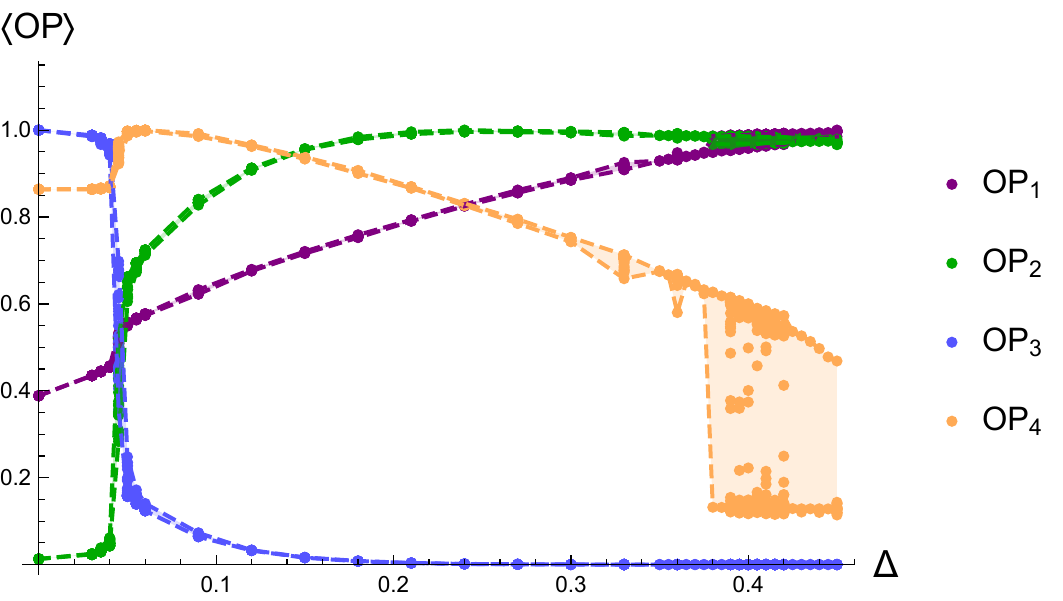}
\caption{\small {Rescaled  order parameters $OP_1$, ..., $OP_4$ in CDT with toroidal spatial topology ($\Sigma = T^3$) measured for many different starting triangulations for each $\Delta$ ($\kappa_0 = 2.0$ is kept fixed), the number of starting configurations being different for various $\Delta$. Each data point denotes $\langle OP \rangle $ measured from last $10^5$ sweeps (or equivalently $10^{12}$ attempted Monte Carlo moves), data from initial thermalization period were skipped. Shaded regions between the dashed lines denote the range of the measured data. Hysteresis is clearly visible for $\Delta \geq 0.38$, especially for the $OP_4$ parameter which is the most sensitive to the bifurcation phase transition.}}
\label{F:Hysteresis}
\end{figure}

Summing up, {the general phase structure of CDT with toroidal spatial topology  is very similar to phase
structure of the CDT with spherical spatial topology and even the positions of phase transition lines are almost the same.  We have observed some qualitative difference in phase transitions but it is impossible to say presently if this implies
that the order of some of the phase transitions should change with topology. It would indeed be somewhat 
surprising if the transitions can be related to continuum physics, in particular UV physics which one 
would imagine was related to short distance phenomena. Short distance phenomena should  be insensitive 
to topology. However, the precise nature of the transitions requires further studies which will be presented 
in forthcoming publications.}

\end{section}


\section*{Acknowledgements}

JGS wishes to acknowledge support of the grant UMO-2016/23/D/ST2/00289 from the National Science Centre, Poland. AG and DN acknowledge support by the National Science Centre, Poland
under grant UMO-2015/17/D/ST2/03479. 
{JA wishes to acknowledge support from the Danish Research Council grant  ``Quantum Geometry''}.
JJ acknowledges the support of the grant UMO-2012/06/A/ST2/00389 from the National Science Centre Poland.



\bibliographystyle{unsrt}
\bibliography{Master}

\end{document}